%% file: main.tex
\documentclass[journal]{IEEEtran}
\usepackage{amsmath,amsfonts}
\usepackage{algorithmic}
\usepackage{array}
\usepackage[caption=false,font=normalsize,labelfont=sf,textfont=sf]{subfig}
\usepackage{textcomp}
\usepackage{stfloats}
\usepackage{url}
\usepackage{verbatim}
\usepackage{graphicx}
\def\BibTeX{{\rm B\kern-.05em{\sc i\kern-.025em b}\kern-.08em
    T\kern-.1667em\lower.7ex\hbox{E}\kern-.125emX}}
\usepackage{balance}

\usepackage{array}  

\usepackage{subfig}

\usepackage{caption}
\usepackage{threeparttable}
\usepackage[table]{xcolor}
\usepackage{multirow}

\usepackage{pifont}
\usepackage{orcidlink}
\usepackage{color}
\definecolor{mitblue}{rgb}{0.88,0.95,0.96}

\usepackage{fontawesome}
\usepackage{xcolor}

\hypersetup{
colorlinks=true,
linkcolor=black,
citecolor=black
}

\usepackage[linesnumbered,ruled]{algorithm2e}

\begin{document}
\title{Channel-Level Mask Guidance for Slimmable Image Compression}

\author{Chen Chen\hspace{-1.5mm}$^{~\orcidlink{0000-0002-8668-5391}}$,  Chenggang Wu\hspace{-1.5mm}$^{~\orcidlink{0000-0002-5952-3426}}$, Yongbiao Chen\hspace{-1.5mm}$^{~\orcidlink{0000-0002-7727-5489}}$,
Weidong Qiu\hspace{-1.5mm}$^{~\orcidlink{0000-0001-6428-1655}}$,~\IEEEmembership{Member,~IEEE,} \\
Zhengwei Qi\hspace{-1.5mm}$^{~\orcidlink{0000-0003-2730-2319}}$,~\IEEEmembership{Member,~IEEE,} and Haibing Guan\hspace{-1.5mm}$^{~\orcidlink{0000-0002-4714-7400}}$,~\IEEEmembership{Senior Member,~IEEE} 

\thanks{Manuscript received XXX XX, 2026. The authors are with the School of Computer Science, Shanghai Jiao Tong University, Shanghai 200240, China (e-mail: chenchen825@sjtu.edu.cn; wuchenggang@sjtu.edu.cn; chenyongbiao0319@sjtu.edu.cn; qiuwd@sjtu.edu.cn; qizhwei@sjtu.edu.cn; hbguan@sjtu.edu.cn). $(Corresponding\ author:\ Zhengwei\ Qi.)$
}
}

\markboth{IEEE TRANSACTIONS ON CIRCUITS AND SYSTEMS FOR VIDEO TECHNOLOGY,~Vol.~18, No.~9, September~2020}%
{Shell \MakeLowercase{\textit{et al.}}: A Sample Article Using IEEEtran.cls for IEEE Journals}

\maketitle

\input{sections/abs_intro}
\input{sections/bg_mt}
\input{sections/elic}
\input{sections/eval}

\section{DISCUSSION}
\label{sec:diss}
Our method focuses on CNN-based VAE image compression. The proposed mask-guided selection and variable-rate modulation can extend to attention and Transformer-based codecs \cite{cheng2020learned, hu2023complexity}. We use uniform scaling ratios within each module; future work may explore non-uniform, operator-level scaling to achieve better balance. Model quantization is promising for reducing computation and cross-platform discrepancies \cite{sun2021learned}. Operator-level redundancy can be further reduced via efficient transposed convolutions and pruned or hardware-parallelized context modeling \cite{chen2026optimizing}. Since entropy coding runs serially on the CPU while other components run on the GPU, it remains the latency bottleneck; future work may explore parallel entropy coding on CPU/GPU \cite{yu2022evaluating, huang2025mans}.

\section{CONCLUSION}
\label{sec:con}

We proposed a channel-level mask-guided slimmable image compression method for flexible complexity and continuous variable-rate control in a single model. A learnable channel-wise mask with sparsity regularization enables non-contiguous top-\(w\) selection, while mask merging and channel reordering convert the model into a standard slimmable network (M2S) without altering the forward pass. Variable-rate control combines \(\lambda\)- and width-conditioned modulation with quantization-step scaling and a progressive freeze-then-fine-tune strategy.

On Kodak, across E2E, Hyperprior, and Checkerboard backbones, M2S consistently outperforms the same-width Slimmable baseline, with the largest BD-rate savings at narrow widths, while reducing storage over independent models. VBR-QW is best on E2E and CKBD, and VBR-QFW on Hyperprior, both approaching the fixed-rate M2S bound. Compared with existing elastic LIC methods, our method offers a better trade-off among rate--distortion performance, parameter efficiency, and latency flexibility. Future work includes hardware-aware entropy coding and video compression.

\section{Acknowledgement}
This work was supported in part by the National Natural Science Foundation of China, and Shanghai Key Laboratory of Scalable Computing and Systems. DeepSeek was used only for language polishing.

\bibliographystyle{IEEEtran}
\bibliography{Ref}

\end{document}

%% file: sections/abs_intro.tex
\begin{abstract}
Learned image compression (LIC) has achieved promising rate-distortion performance, but most methods lack scalability for diverse platforms with varying computational and storage constraints. Existing slimmable LIC methods achieve flexible complexity via consecutive-channel activation, but suffer from two critical limitations. First, forced consecutive-channel selection degrades narrow-model performance at low bitrates. Second, the lack of stable variable-rate control limits their adaptability to varying bandwidth conditions.

To address these limitations, we propose a channel-level mask-guided slimmable image compression method. A learnable channel-wise mask with sparsity-promoting regularization enables non-contiguous top-$w$ channel selection, allowing narrow models to use the most informative channels. After training, mask merging and channel reordering convert the model into a standard slimmable network via Mask-to-Slimmable (M2S) conversion, supporting efficient inference via channel pruning without architectural changes. For variable-rate control, we combine $\lambda$- and width-conditioned modulation with quantization-step scaling and a progressive training pipeline stabilized by a freeze-then-fine-tune strategy.

Experiments on Kodak across E2E, Hyperprior, and Checkerboard backbones show that our method achieves superior rate-distortion performance across multiple widths and bitrates, with significant gains for narrow models at low bitrates. At the narrowest width ($w=48$), it achieves up to 19.22\%, 24.11\%, and 22.76\% BD-rate savings over the same-width Slimmable baseline on the E2E, Hyperprior, and CKBD backbones, respectively. A single VBR model replaces 40 independent models, reducing storage by $14.8\times$--$15.9\times$. Compared with existing elastic LIC methods, our method outperforms the same-width Slimmable baseline and compares favorably in terms of rate-distortion performance, parameter efficiency, and latency flexibility.
The code is available at https://github.com/sjtu-tcloud/MGSIC.
\end{abstract}

\begin{IEEEkeywords}
Neural Networks, Image Compression, Scalable Coding, Variable Bitrate.
\end{IEEEkeywords}

\section{INTRODUCTION}
\label{sec:intro}
\IEEEPARstart{L}{earned} image compression (LIC) has achieved remarkable progress in rate-distortion performance, even surpassing traditional codecs (e.g., H.265/H.266 intra-mode) in certain settings \cite{balle2017end, balle2018variational, minnen2018joint, cheng2020learned, liu2023learned, guoevc}. However, efficient compression for resource-constrained scenarios remains underexplored \cite{minnen2020channel, he2021checkerboard, he2022elic, yang2021slimmable}. Traditional scalable coding (e.g., SVC/SHVC) transmits layered bitstreams but suffers from high decoding complexity \cite{schwarz2007overview, boyce2015overview}. Several elastic LIC methods have thus been developed for quality and spatial scalability \cite{guoevc, yang2021slimmable, mei2021learning, zhang2023exploring, guo2023cbanet, zhang2023elfic, li2025one}.

\begin{figure}[t!]
\centering
  \setlength{\abovecaptionskip}{0.1cm}
  \setlength{\belowcaptionskip}{-0.6cm}
  \includegraphics[width=0.5\textwidth]{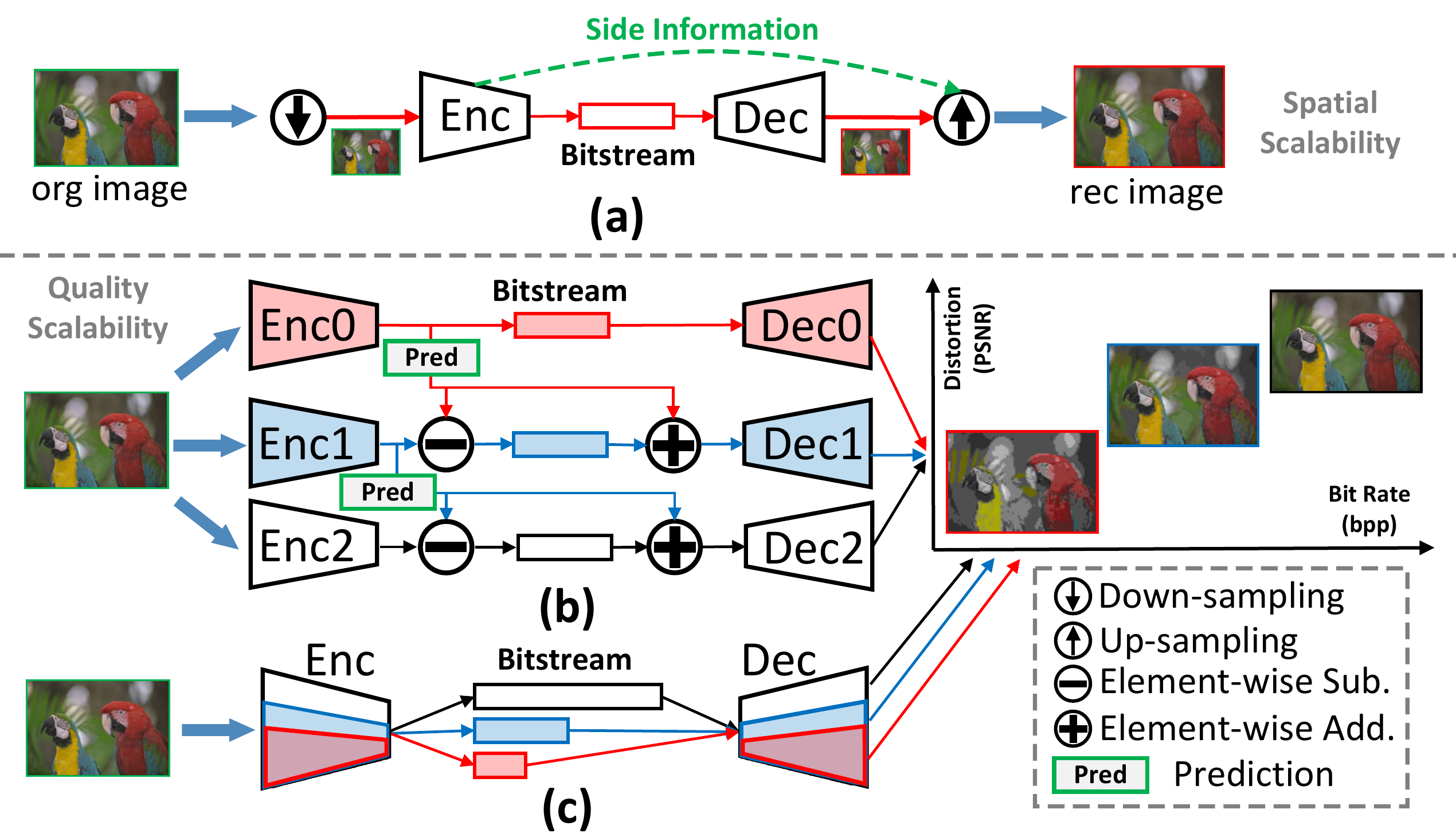}
  \caption{Three typical styles of Elastic LIC methods. (a) Down-Up Sampling Style. (b) Scalable Style. (c) Slimmable Style. Quantization and entropy coding are omitted. Element-wise addition could also be replaced by channel-wise concatenation.}
  \label{fig:1}
\end{figure}

As shown in Fig.~\ref{fig:1}, elastic LIC methods fall into two categories: spatial scalability and quality scalability. Spatial scalability downsamples the input at the encoder and upsamples the decoded frame at the decoder~\cite{zamfir2024see, li2017convolutional, park2023compass, shen2011down}, often incorporating side information to enhance quality~\cite{park2023compass, lee2022dpict}. Quality scalability further splits into learning-based scalable coding~\cite{mei2021learning, zhang2023exploring, guoevc, dasari2022swift} and slimmable coding~\cite{yang2021slimmable, guo2023cbanet, zhang2023elfic, li2025one, hu2023complexity}. Among these styles, slimmable coding offers the best flexibility-efficiency trade-off, yet suffers from two critical limitations. First, forced consecutive-channel selection severely degrades narrow-model performance at low bitrates. Second, existing solutions lack effective variable-rate control, limiting their adaptability to varying bandwidth conditions.

Variable-rate control is essential for practical codecs. Existing approaches typically introduce modulation networks conditioned on $\lambda$ \cite{tong2023qvrf, yang2020variable, choi2019variable}, but they mostly target fixed-width models. Integrating them into a slimmable framework is nontrivial: naive $\lambda$-sampling fails to decouple width adaptation from rate modulation, causing training instability.

To address these challenges, we propose a channel-level mask-guided slimmable image compression method. A learnable structured-sparsity mask enables non-contiguous top-\(w\) selection so that narrow models retain the most informative channels. Mask merging and channel reordering then convert the mask model into a standard slimmable network (M2S) for channel-pruned inference without architectural changes. For variable-rate control, we combine \(\lambda\)- and width-conditioned modulation with quantization-step scaling and a progressive freeze-then-fine-tune strategy to decouple width adaptation from rate modulation.

Our main contributions are summarized as follows:
\begin{itemize}
\item We propose a learnable channel-wise mask with sparsity regularization for non-contiguous top-\(w\) selection, improving narrow-model performance at low bitrates.
\item We design an M2S conversion (mask merging and channel reordering) that transforms the mask model into slimmable model without altering the forward pass, making active channels contiguous for efficient inference.
\item We develop a variable-rate framework with \(\lambda\)- and width-conditioned modulation, quantization-step scaling, and a progressive freeze-then-fine-tune strategy, enabling continuous rate control in a single slimmable model.
\item Experiments across E2E, Hyperprior, and Checkerboard backbones show up to 19.22\%, 24.11\%, and 22.76\% BD-rate savings over the same-width Slimmable baseline at \(w=48\), with comprehensive comparisons in parameter efficiency, latency, and traditional codecs.
\end{itemize}

%% file: sections/bg_mt.tex
\section{BACKGROUND AND MOTIVATION}
\label{sec:bg}

\subsection{Learned Image Compression}

Three representative LIC models are reviewed: E2E model \cite{balle2017end}, Hyperprior model \cite{balle2018variational}, and Context model \cite{minnen2018joint}.

\begin{figure}[t!]
\centering
  \setlength{\abovecaptionskip}{0.1cm}
  \setlength{\belowcaptionskip}{-0.5cm}
  \includegraphics[width=0.5\textwidth]{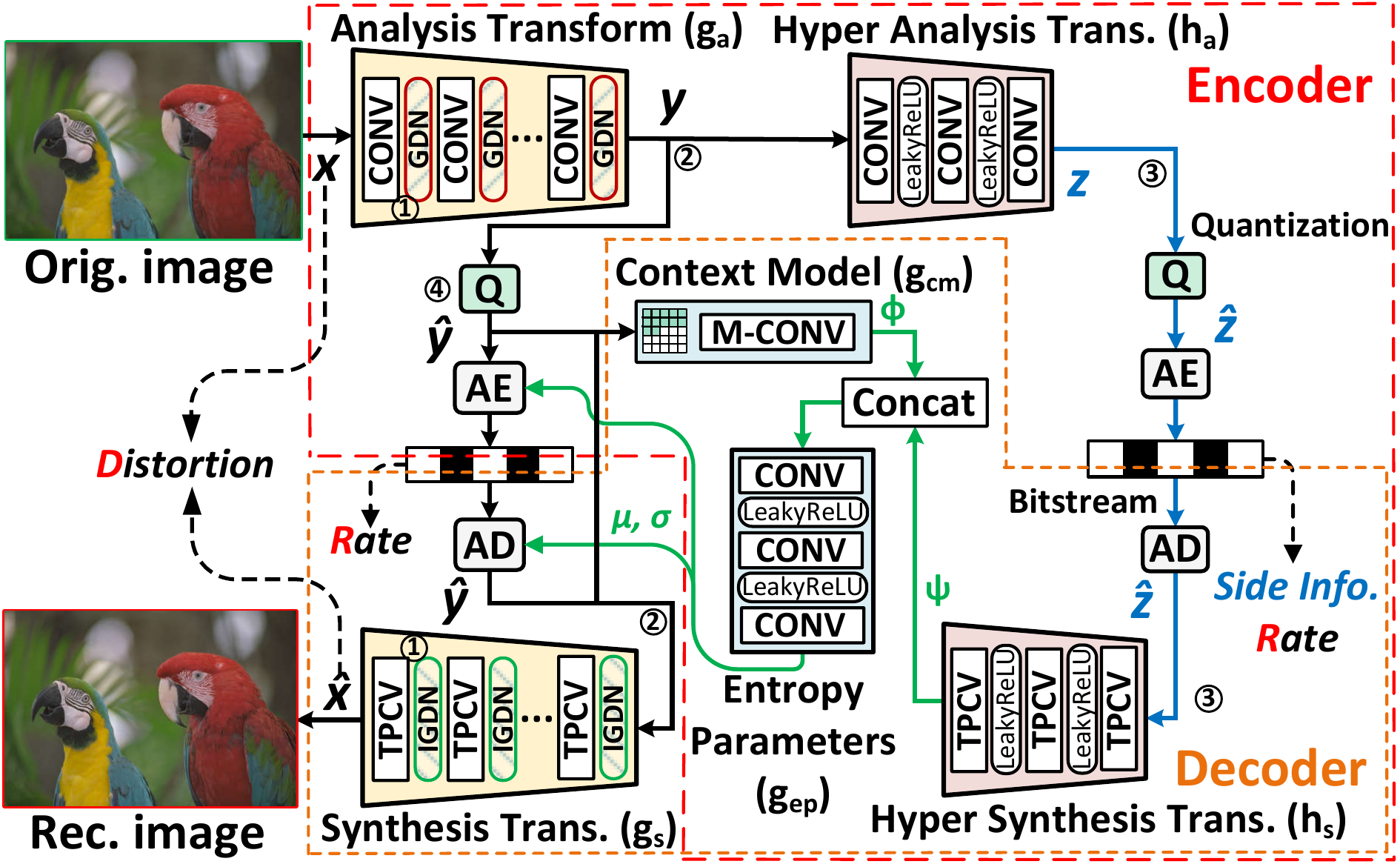}
  \caption{Overview of typical LIC model with context model. \(g_a\) and \(g_s\) consist of Convolution (CONV) / Transposed Convolution (TPCV) with Generalized Divisive Normalization (GDN) / Inverse GDN (IGDN), respectively. ``AE/AD'' denote arithmetic en-/de-coding. ``Concat'' denotes channel-wise tensor concatenation. ``Q'' denotes quantization. The numbers \ding{172}--\ding{175} denote the four positions for variable-rate methods summarized in Table~\ref{tab:1}: \ding{172} feature maps, \ding{173} latent \(y\), \ding{174} side information \(z\), and \ding{175} the quantizer.}
  \label{fig:2}
\end{figure}

\noindent\textbf{E2E Model:} an end-to-end optimized model is proposed with learnable nonlinear analysis/synthesis transforms \(g_a/g_s\) and arithmetic coding, jointly optimizing rate and distortion \cite{balle2017end}. As shown in Fig.~\ref{fig:2}, input \(x \in \mathbb{R}^{H\times W\times 3}\) is converted into latent \(y\) by \(g_a\). Quantization (Q) casts \(y\) into integer \(\hat{y}\), which is compressed by arithmetic encoding (AE). Arithmetic decoding (AD) restores \(\hat{y}\), and \(\hat{x}\) is reconstructed by \(g_s\):
\begin{equation}
\label{eq:e2e}
    y = g_a(x),\quad
    \hat{y} = Q(y),\quad
    \hat{x} = g_s(\hat{y})
\end{equation}
Usually, uniform noise is used for quantization during training, and hard rounding during inference. The loss function is:
\begin{equation}
\label{eq:e2e_loss}
    \mathcal{L} = R + \lambda D
\end{equation}
where \(R\) is the bitrate (bit per pixel, bpp), \(D\) is the distortion (e.g., MSE), and \(\lambda\) controls the rate-distortion trade-off. Its factorized entropy model (FEM) assumes i.i.d. elements within each channel, so it learns one distribution per channel and requires no per-element \(\mu\) or \(\sigma\).

\noindent\textbf{Hyperprior Model:} the hyperprior architecture is introduced, adding a side branch to capture spatial dependencies in \(y\) \cite{balle2018variational}. As shown in Fig.~\ref{fig:2}, \(y\) undergoes hyper analysis/synthesis transforms \(h_a/h_s\) to estimate per-element mean \(\mu\) and scale \(\sigma\) for \(y\), forming a Gaussian conditional entropy model. The side information \(z\) between \(h_a\) and \(h_s\) is quantized and compressed:
\begin{equation}
    z = h_a(y),\quad
    \hat{z} = Q(z),\quad
    (\mu,\sigma) = h_s(\hat{z})
\end{equation}
The loss extends Eq.~\ref{eq:e2e_loss} with a side-information rate term:
\begin{equation}
\label{eq:hyperprior_loss}
    \mathcal{L} = R + R_{\text{side}} + \lambda D
\end{equation}
where \(R_{\text{side}}\) is the bitrate of \(\hat{z}\).

\noindent\textbf{Context Model:} Minnen \textit{et al.} \cite{minnen2018joint} proposed a context model with autoregressive priors. Its context module \(g_{cm}\) is usually a masked convolution:
\begin{equation}
    g_{cm}(x) = (M_{\mathrm{ctx}} \odot \mathbf{W}) \times x + b
\end{equation}
where \(\odot\) is the Hadamard product and 0-1 mask $M_{\mathrm{ctx}}$ ensures only decoded pixels (left and above in Z-order) are visible.

For pixel \(i\), \(g_{cm}\) generates context feature \(\phi_i\) from visible \(\hat{y}_{<i}\). It is concatenated with hyperprior feature \(\psi_i\) from \(h_s(\hat{z})\), and fed into entropy parameter module \(g_{ep}\):
\begin{equation}
\begin{gathered}
\phi_i = g_{cm}(\hat{y}_{<i}),\quad
\psi_i = h_s(\hat{z})_i,\quad
(\mu,\sigma)_i = g_{ep}(\psi_i,\phi_i)
\end{gathered}
\end{equation}
where \(i\) indexes the spatial position in \(\hat{y} \in \mathbb{R}^{C_{y} \times H_{y} \times W_{y}}\), with \(C_{y}\), \(H_{y}\), and \(W_{y}\) denoting channel number, height, and width of \(\hat{y}\). \(\hat{y}_{<i}\) denotes encoded/decoded pixels.

Subsequent studies accelerated context modeling by partitioning $y$ either spatially (e.g., Checkerboard \cite{he2021checkerboard}) or channel-wise (e.g., \cite{minnen2020channel}), enabling parallel prediction. We adopt E2E, hyperprior, and Checkerboard models as backbones.

\subsection{Slimmable Image Compression}
\label{sec:sic}

Yu \textit{et al.}~\cite{yu2019slimmable,yu2019universally} proposed slimmable neural networks that support inference at different widths without retraining; Yang \textit{et al.}~\cite{yang2021slimmable} introduced this paradigm to LIC with SlimCAE, which jointly trains multiple width sub-networks, each assigned its own RD trade-off \(\lambda_w\) for the target width. For target widths \(\mathcal{W}=\{w_1,w_2,\ldots,w_m\}\), the objective is:
\begin{equation}
\label{eq:slimmable_loss}
    \mathcal{L}_{\text{slim}} = \sum_{w \in \mathcal{W}} \left( R_w + \lambda_w D_w \right)
\end{equation}
where \(R_w\) and \(D_w\) are the bitrate and distortion at width \(w\), and \(\lambda_w\) is the width-specific Lagrangian multiplier.

\begin{figure*}[t!]
\centering
  \setlength{\abovecaptionskip}{0.1cm}
  \setlength{\belowcaptionskip}{-0.2cm}
  \includegraphics[width=0.98\textwidth]{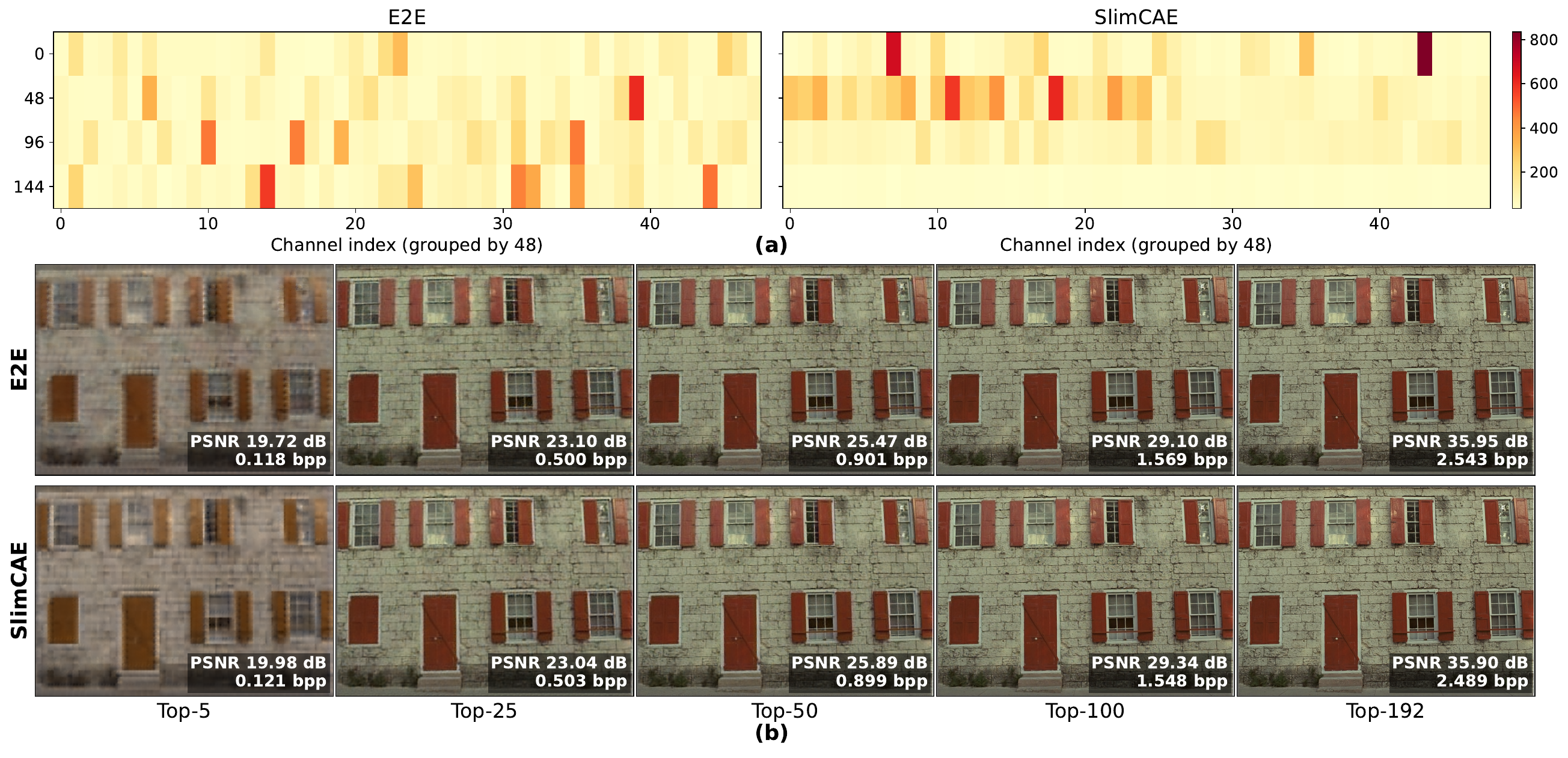}
  \caption{Channel importance and its effect on reconstruction ($\lambda=0.18,w=192$).
  (a) Channel-wise $\ell_2$ normalization of the latent $y$, where each row represents 48 consecutive channels.
  (b) Joint reconstructions using only the top-$K$ channels ranked by $\ell_2$ norm, with $K \in \{5, 25, 50, 100, 192\}$. PSNR and bpp are annotated on each reconstruction.
  }
  \label{fig:channel_priority}
\end{figure*}

Slimmable methods define sub-networks as the first \(K\) consecutive channels, assuming monotonic importance in index order. Although this consecutive-prefix structure is system-friendly (active weights and activations form contiguous memory blocks, so width adaptation maps to contiguous tensor slices and MAC units without dynamic gather or scatter), training enforces only coarse group-level ordering, so narrow models may inherit suboptimal channels. Fig.~\ref{fig:channel_priority}(a) shows scattered high-response channels in E2E, whereas SlimCAE shows clear gaps between channel groups; even within the first 48-channel group, some channels have lower \(\ell_2\) normalization than later ones. A fixed prefix thus includes less informative channels and excludes more useful ones. Fig.~\ref{fig:channel_priority}(b) shows top-5 channels already capture the dominant low-frequency layout, with PSNR increasing monotonically with \(K\), confirming a small informative subset suffices for faithful reconstruction. This motivates non-contiguous selection: at low bitrates, image energy concentrates in a few channels, so a fixed prefix includes redundant channels and excludes more informative ones. This effect diminishes as width increases.

\subsection{Variable Rate and Complexity Trade-off}
\label{sec:var-rate}

\begin{table}[t]
\setlength{\abovecaptionskip}{0.1cm}
\caption{Summary of variable bitrate research.}
\centering
\renewcommand{\arraystretch}{1.15}
\begin{tabular}{m{1.8cm}|m{0.7cm}|m{5.0cm}}
\hline
\hline
& \textbf{Pos.\(^{a}\)} & \multicolumn{1}{c}{\textbf{Training settings\(^{b}\)}} \\ \hline
\textbf{AG-VAE} \cite{cui2021asymmetric} & \ding{173} & Random training \\ \hline
\textbf{DCVC} \cite{li2022hybrid} & \ding{173} & Sequential training \\ \hline
\textbf{CAE} \cite{choi2019variable} & \ding{172}\ding{175} & Sequential training with bin size=1; random bin size \\ \hline
\textbf{QVRF} \cite{tong2023qvrf} & \ding{173} & Train with \(\lambda_{\max}\), then sequential training with other \(\lambda\)s \\ \hline
\textbf{DCT-RES} \cite{duong2023multi} & \ding{172} & Random training, biased toward larger \(\lambda\) \\ \hline
\textbf{MAE} \cite{yang2020variable} & \ding{172} & Sequential training with total loss of all \(\lambda\)s \\ \hline
\textbf{WPA} \cite{wang2024variable} & \ding{172}\ding{173}\ding{174} & Random training, biased toward larger \(\lambda\) \\ \hline
\textbf{DM} \cite{lin2021deeply} & \ding{172}\ding{173} & Sequential training \\ \hline
\textbf{SF} \cite{chen2020variable} & \ding{173}\ding{174} & Train with \(\lambda_{\max}\) (freeze except scaling factors), then sequential training \\
\hline
\hline
\end{tabular}
\begin{tablenotes}
    \item
    \(^{a}\) Positions \ding{172}--\ding{175} correspond to Fig.~\ref{fig:2}.
    \(^{b}\) Random: \(\lambda\sim\mathrm{Unif}(\Lambda)\); Sequential: \(\lambda\) in ascending order over \(\Lambda\); ``biased toward larger \(\lambda\)'': larger \(\lambda\) sampled more often.
\end{tablenotes}
\label{tab:1}
\end{table}

Variable-rate methods adapt a single model to varying RD trade-offs. JPEG scales quantization tables, whereas standard LIC trains a separate network per \(\lambda\)~\cite{balle2017end}. Recent methods condition a single model on \(\lambda\) or bin size~\cite{tong2023qvrf,yang2020variable,choi2019variable,cui2021asymmetric,li2022hybrid,duong2023multi,wang2024variable,lin2021deeply,chen2020variable} at four locations (Fig.~\ref{fig:2}): \ding{172} feature maps, \ding{173} latent \(y\), \ding{174} latent \(z\), and \ding{175} the quantizer. 

Table~\ref{tab:1} summarizes representative methods. QVRF~\cite{tong2023qvrf} and SF~\cite{chen2020variable} use fixed \(\lambda_{\max}\), while others sample larger \(\lambda\) more often~\cite{duong2023multi,wang2024variable}. AG-VAE~\cite{cui2021asymmetric} and DCVC~\cite{li2022hybrid} scale only latent \(y\), whereas MAE~\cite{yang2020variable} and DM~\cite{lin2021deeply} modulate intermediate features. QVRF uses a scalar regulator, while others use channel-wise vectors. SF freezes the main model and updates only scaling factors, degrading low-bitrate performance.

Existing variable-rate methods target fixed-width models, and integrating them into a slimmable framework couples width and rate training, causing instability. This is because \(\lambda\) and width are intrinsically coupled: larger \(\lambda\) requires wider models, while narrow models saturate at high bitrates \cite{balle2018variational,cheng2020learned}. Since low bitrates also cause widths to converge, we decouple width adaptation from rate modulation.

\subsection{Structured Pruning and Mask-based Channel Selection}

Structured pruning removes channels for standard-hardware deployment \cite{hossain2024structured}, while dynamic pruning prunes activations per input but requires customized hardware \cite{yin2022universal}. Yin \textit{et al.}~\cite{yin2022structural} learn binary channel-importance masks and prune at inference; EVC~\cite{guoevc} sparsifies learnable masks between convolutions and merges them into adjacent layers. However, these methods target fixed-width or static pruning and do not support non-contiguous selection at target widths or joint width/rate adaptation, motivating our mask-guided slimmable framework.

%% file: sections/elic.tex
\section{DESIGN METHODOLOGY}
\label{sec:design}

\begin{figure*}[t!]
\centering
  \setlength{\abovecaptionskip}{0.1cm}
  \setlength{\belowcaptionskip}{-0.5cm}
  \includegraphics[width=0.99\textwidth]{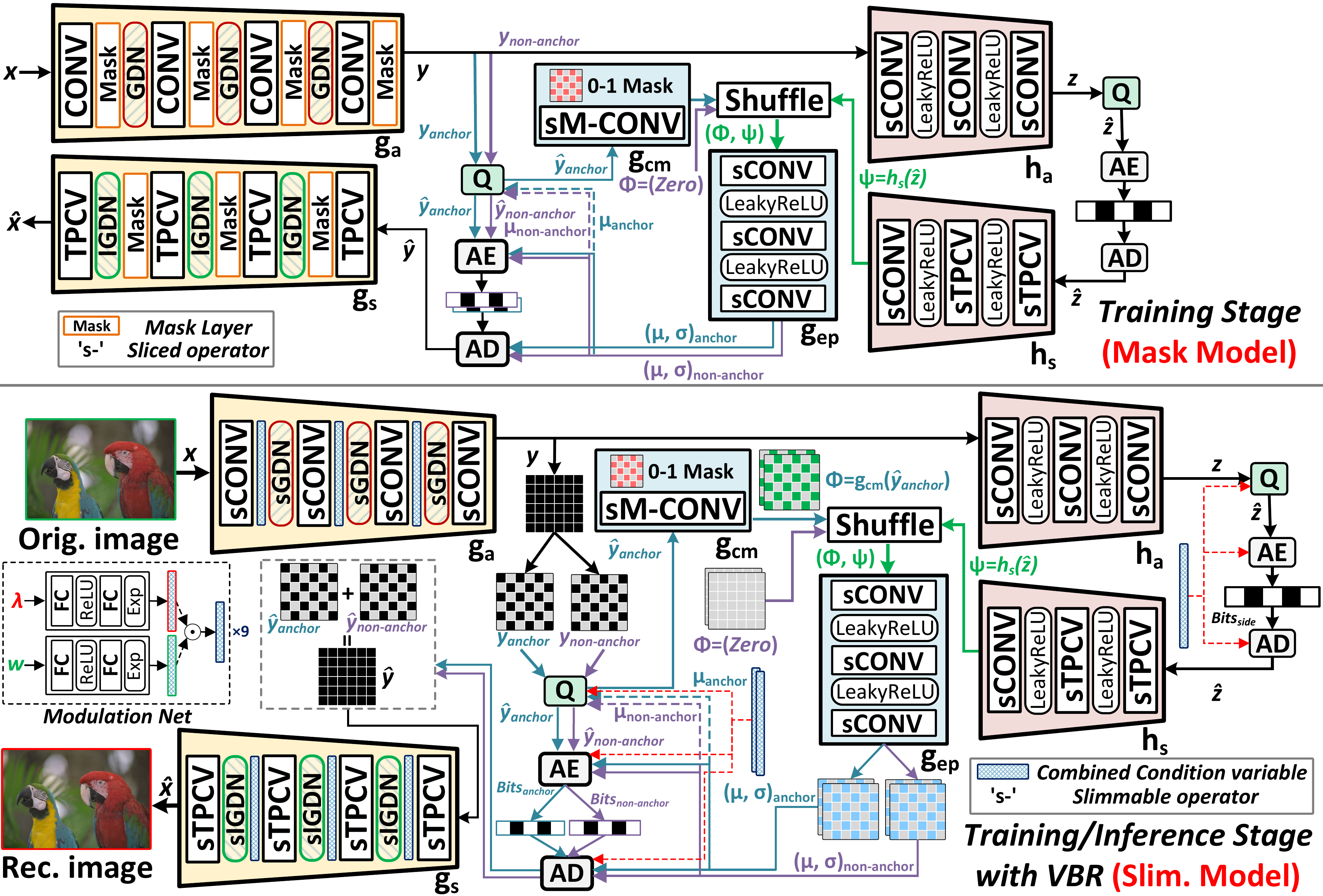}
  \caption{Overview of the proposed method. (Top) Training stage of the Mask model with learnable masks and sliced operators. (Bottom) Training and inference stage of the Slimmable model with variable-rate control, where \(\lambda\)- and width-conditioned modulation networks generate combined variables via Hadamard product, applied to \(g_a/g_s\) features and quantization step sizes for \(y/z\), and channel shuffling aligns \((\Phi,\Psi)\) across widths. The M2S conversion between the two stages is detailed in Fig.~\ref{fig:reordering}. Operator configurations are in the supplementary material.}
  \label{fig:architecture}
\end{figure*}

Our design is motivated by two observations. First, the lottery ticket hypothesis \cite{franklelott2018lottery} suggests that channel importance learned at maximum width generalizes across widths. Second, in transform coding, natural image energy concentrates in low frequencies, making channel importance largely consistent across width budgets. We therefore learn a universal channel ranking at maximum width to guide non-contiguous selection.

\subsection{Overall Architecture}
Fig.~\ref{fig:architecture} shows the pipeline with two stages connected by an M2S conversion.
\textbf{(1) Mask training:} a Mask model is trained with learnable masks and sliced operators (Sec.~\ref{sec:ops}). The masks evaluate channel importance and enable non-contiguous selection. The transforms \(g_a/g_s\) contain mask layers, while \(h_a\), \(h_s\), \(g_{cm}\), and \(g_{ep}\) use sliced operators, and Concat is replaced by channel shuffling (Sec.~\ref{sec:channel_shuffle}).
\textbf{(2) M2S conversion:} mask merging and channel reordering convert the Mask model into a standard Slimmable network without altering the forward pass (Fig.~\ref{fig:reordering}; Sec.~\ref{sec:reordering}). This allows fine-tuning to mitigate the sparsity regularization cost and makes active channels contiguous; modules with only sliced operators are reordered by their adjacent mask layers.
\textbf{(3) VBR training and inference:} the converted model is fine-tuned and deployed for variable-rate and width-adaptive compression (Sec.~\ref{sec:modulation}). The \(\lambda\)- and width-conditioned modulation networks generate combined variables via Hadamard product, applied to \(g_a/g_s\) features and quantization step sizes for \(y\) and \(z\), and channel shuffling aligns \((\Phi,\Psi)\) across widths. A progressive multi-stage strategy enables one model to cover multiple rate-distortion trade-offs.

\subsection{Channel-Level Mask and Sliced Operators}
\label{sec:ops}

\begin{figure*}[t!]
\centering
  \setlength{\abovecaptionskip}{0.1cm}
  \setlength{\belowcaptionskip}{-0.5cm}
  \includegraphics[width=0.99\textwidth]{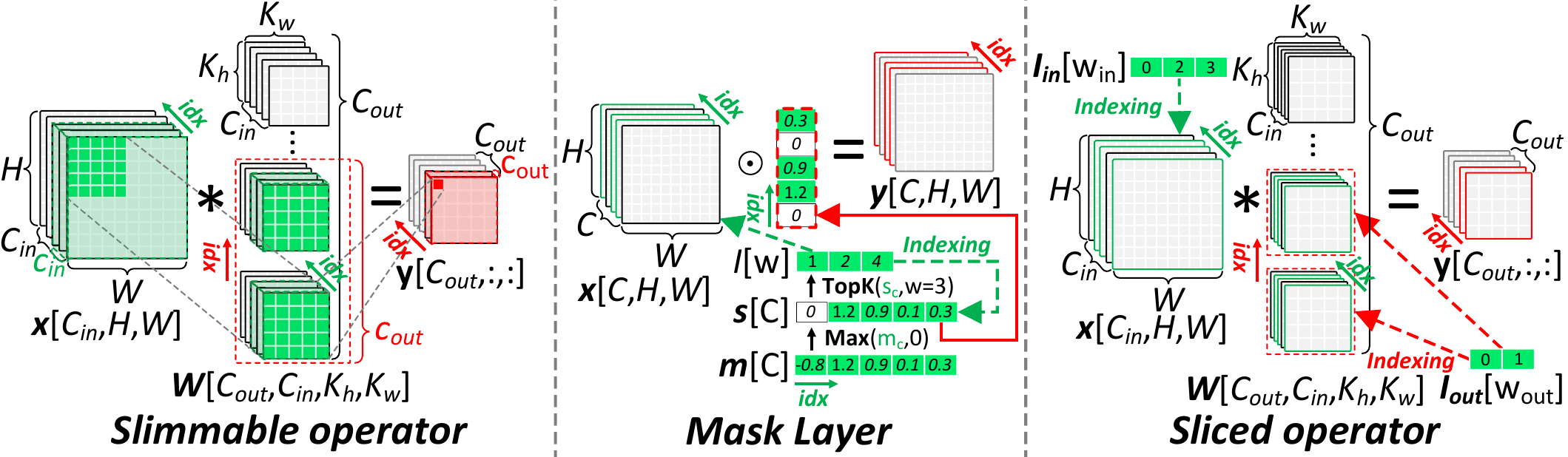}
  \caption{Three channel selection strategies. (a) Traditional slimmable operator with a fixed prefix (\(w_{\mathrm{in/out}}=3/2\)). (b) Mask layer selecting top-\(w\) channels ($w=3$) via a learnable mask. (c) Sliced operator with index sets $\mathcal{I}_{\mathrm{in}}=[0,2,3]$, $\mathcal{I}_{\mathrm{out}}=[0,1]$, decoupling selection from execution. Gray borders in $y$ indicate zero-padded data.}
  \label{fig:operators}
\end{figure*}

Fig.~\ref{fig:operators} illustrates three channel selection strategies: (1) the slimmable operator using a fixed prefix of consecutive channels, (2) the learnable mask layer selecting non-contiguous channels, and (3) the sliced operator computing on the non-contiguous channel subsets selected by the learned mask.

\subsubsection{Slimmable Operator}
As shown in Fig.~\ref{fig:operators}(a), slimmable operator is executed on a sliced subset of channels, strictly constrained to consecutive channels starting from index 0 \cite{yang2021slimmable, yu2019slimmable, yu2019universally}. Let \(w_{\mathrm{in}}\) and \(w_{\mathrm{out}}\) be the active input and output widths. The input \(x \in \mathbb{R}^{C_{\mathrm{in}} \times H \times W}\) is truncated to \(x[0:w_{\mathrm{in}}, :, :]\), and the weight \(\mathbf{W} \in \mathbb{R}^{C_{\mathrm{out}} \times C_{\mathrm{in}} \times K_h \times K_w}\) and bias \(b \in \mathbb{R}^{C_{\mathrm{out}}}\) are sliced accordingly:
\begin{equation}
y[:w_{\mathrm{out}}, ::] = \mathbf{W}[:w_{\mathrm{out}}, :w_{\mathrm{in}}, ::] \star x[:w_{\mathrm{in}}, ::] + b[:w_{\mathrm{out}}],
\end{equation}
where \(\star\) denotes convolution (CONV). The output is zero-padded to the full dimension if necessary.

The same channel slicing applies to other operators: TPCV slices its weight and bias identically, channel-wise normalization layers such as GDN/IGDN slice their per-channel parameters, and parameter-free operators such as LeakyReLU act element-wise and need no modification.

\subsubsection{Mask Layer}

We propose a learnable mask layer that identifies the most informative channels for each width, as illustrated in Fig.~\ref{fig:operators}(b). It is inserted after selected CONV/TPCVs in \(g_a\) and \(g_s\) (Fig.~\ref{fig:architecture}). Inspired by the mask decay in EVC~\cite{guoevc}, we assign a trainable scalar \(m_c\) to each channel, initialized to \(m_c = 1.05\), forming a mask vector \(\mathbf{m} \in \mathbb{R}^C\). The mask score is \(s_c = \max(m_c, 0)\). Given a target width \(w\), the mask layer selects the top-\(w\) channels \(\mathcal{I} = \mathrm{TopK}(\mathbf{s}, w)\) and produces
\begin{equation}
x_{\mathrm{out}}[c, :, :] =
\begin{cases}
x[c, :, :] \cdot s_c, & c \in \mathcal{I},\\
0, & \text{otherwise}.
\end{cases}
\label{eq:mask_forward}
\end{equation}
Since \(\mathcal{I}\) is obtained by ranking the same scores \(s_c\), the selected sets are nested, i.e., \(\mathcal{I}_{w_1} \subset \mathcal{I}_{w_2}\) for \(w_1 < w_2\), which ensures that narrower sub-networks reuse the channels of wider ones and facilitates the subsequent M2S conversion.

\begin{figure}[t!]
\centering
\setlength{\abovecaptionskip}{0.1cm}
\setlength{\belowcaptionskip}{-0.4cm}
\includegraphics[width=0.48\textwidth]{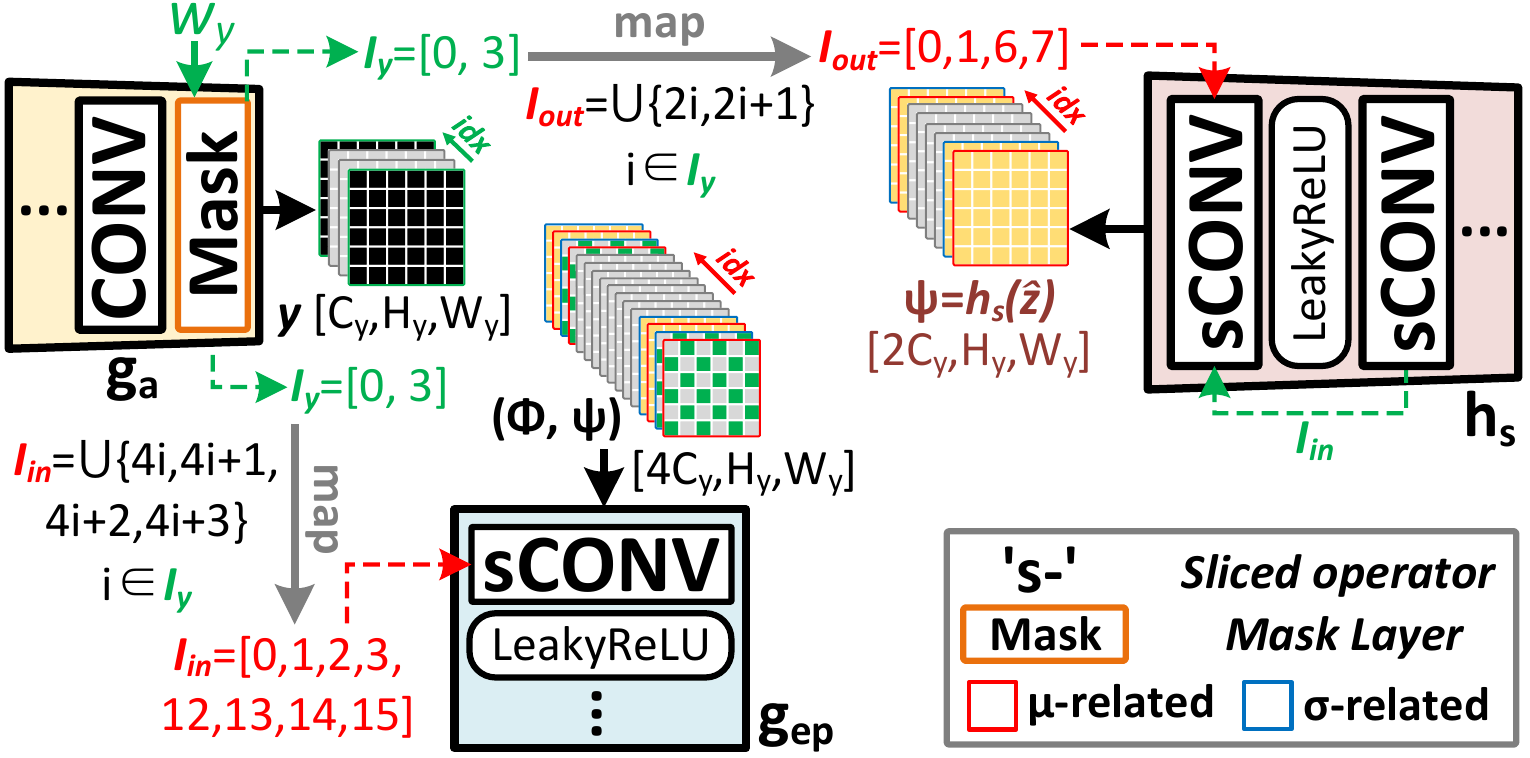}
\caption{Channel index propagation from \(\hat{y}\) to \(h_s\) and \(g_{ep}\). The selected channels \(\mathcal{I}_y\) map to \(\mathcal{I}_{\mathrm{out}}^{hs} = \bigcup_{i \in \mathcal{I}_y}\{2i, 2i+1\}\) in \(h_s\) and to \(\mathcal{I}_{\mathrm{in}}^{ep} = \bigcup_{i \in \mathcal{I}_y}\{4i, 4i+1, 4i+2, 4i+3\}\) in \(g_{ep}\). Unselected channels are zero-padded (gray).}
\label{fig:index_propagation}
\end{figure}

During training, we inject a sparsity-promoting gradient into each mask parameter through a backward hook:
\begin{equation}
\frac{\partial \mathcal{L}}{\partial m_c} \leftarrow \frac{\partial \mathcal{L}}{\partial m_c} + \gamma \cdot |m_c - 1|,
\label{eq:mask_grad}
\end{equation}
which is equivalent to penalizing
\begin{equation}
\mathcal{L}_{\mathrm{sparse}} = \frac{\gamma}{2} \sum_c \,(m_c - 1)\, |m_c - 1|.
\label{eq:mask_loss}
\end{equation}
Starting from \(m_c = 1.05\), this penalty pushes \(m_c\) toward \(1\) and then toward \(0\), pruning unimportant channels. Since an overly strong decay degrades high-bitrate reconstruction, we schedule \(\gamma = \gamma_0 / \lambda\), where \(\gamma_0\) is a preset coefficient (e.g., \(1 \times 10^{-5}\)). This applies stronger sparsification at low bitrates while relaxing it at high bitrates.

Mask layers are inserted only after CONV/TPCVs in \(g_{a}\) and \(g_{s}\), keeping a constant maximum width during training via full-dimensional weights with unselected channels zeroed. Since \(m_c\) stays positive and the mask is merged into the preceding convolution at M2S conversion (Sec.~\ref{sec:reordering}), \(s_c=0\) does not arise at inference.

\subsubsection{Sliced Operators}
\label{sec:sliced}

Unlike slimmable operators constrained to a fixed prefix, sliced operators act on non-contiguous channel subsets selected by the learned mask. For input \(x \in \mathbb{R}^{C_{\mathrm{in}} \times H \times W}\), weight \(\mathbf{W} \in \mathbb{R}^{C_{\mathrm{out}} \times C_{\mathrm{in}} \times K_h \times K_w}\), and bias \(b \in \mathbb{R}^{C_{\mathrm{out}}}\), the sliced operation is
\begin{equation}
y[\mathcal{I}_{\mathrm{out}}, :, :] = \mathbf{W}[\mathcal{I}_{\mathrm{out}}, \mathcal{I}_{\mathrm{in}}] \star x[\mathcal{I}_{\mathrm{in}}, :, :] + b[\mathcal{I}_{\mathrm{out}}],
\label{eq:sliced}
\end{equation}
where \(\star\) denotes convolution and unselected output channels are zero-padded.

The entropy-estimation modules \(h_a\), \(h_s\), \(g_{cm}\), and \(g_{ep}\) contain no Mask layer. Instead, they use sliced operators to enable the channel groups corresponding to the latent set \(\mathcal{I}_y\) selected by the preceding Mask layer in \(g_a\), ensuring that the predicted \(\mu/\sigma\) data remain aligned with the latent elements (Fig.~\ref{fig:index_propagation}). Specifically, \(h_a\) operates directly on \(\mathcal{I}_y\); \(g_{cm}\) takes \(\hat{y}\) restricted to \(\mathcal{I}_y\) as input and outputs paired context features indexed by \(\mathcal{I}_{\mathrm{out}}^{hs}\); \(h_s\) progressively maps \(\mathcal{I}_y\) to \(\mathcal{I}_{\mathrm{out}}^{hs}\) so that each active latent channel yields a paired \((\mu,\sigma)\); and \(g_{ep}\) receives the four entropy-related features per active channel, with input indices \(\mathcal{I}_{\mathrm{in}}^{ep}\) derived in Sec.~\ref{sec:channel_shuffle}. Hence, all entropy-related modules operate on \(w_y = |\mathcal{I}_y|\) active latent channels.

\noindent\textbf{Index propagation.}
To keep each active latent channel paired with its mean and scale, the last sCONV of \(h_s\) outputs
\begin{equation}
\mathcal{I}_{\mathrm{out}}^{hs} = \bigcup_{i \in \mathcal{I}_y} \{ 2i,\; 2i+1 \},
\label{eq:index_hs}
\end{equation}
where \(2i\) and \(2i+1\) correspond to the mean and scale of the \(i\)-th active channel. For \(\mathcal{I}_y = \{0, 3\}\), this yields \(\mathcal{I}_{\mathrm{out}}^{hs} = \{0, 1, 6, 7\}\). The corresponding input index set for \(g_{ep}\), \(\mathcal{I}_{\mathrm{in}}^{ep} = \bigcup_{i \in \mathcal{I}_y}\{4i, 4i+1, 4i+2, 4i+3\}\), is derived in Sec.~\ref{sec:channel_shuffle}, together with the explicit channel shuffling operator used in the Slimmable model.

\subsection{Channel Shuffling}
\label{sec:channel_shuffle}

\begin{figure}[t!]
\centering
  \setlength{\abovecaptionskip}{0.1cm}
  \setlength{\belowcaptionskip}{-0.4cm}
  \includegraphics[width=0.48\textwidth]{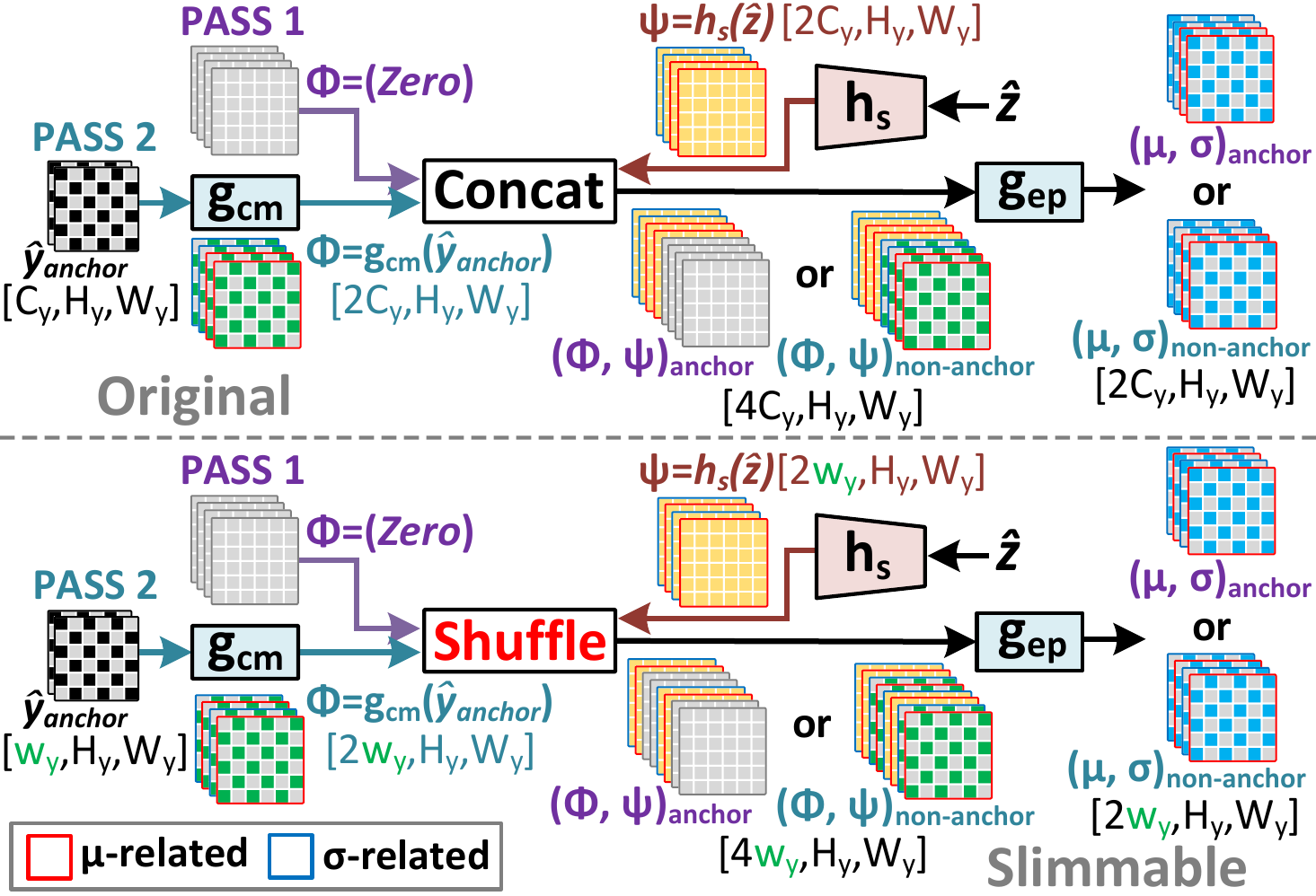}
\caption{Checkerboard model for predicting \((\mu, \sigma)\) of \(\hat{y}\), and its slimmable adaptation. (top) Original two-pass prediction: Pass 1 uses only \(\Psi\); Pass 2 uses \(\Psi\) and \(\Phi\) from decoded anchors. (bottom) Slimmable Model with contiguous active channels and channel shuffling. In Mask Model, channels may be non-contiguous and are handled by sliced operators with set \(\mathcal{I}_y\).}
\label{fig:shuffle}
\end{figure}

In the Checkerboard model \cite{he2021checkerboard}, the latent \(y \in \mathbb{R}^{C_y \times H_y \times W_y}\) is split into anchor and non-anchor sets. As shown in Fig.~\ref{fig:shuffle}(top), \((\mu, \sigma)\) prediction proceeds in two passes: Pass 1 predicts anchors using only the hyperprior feature \(\Psi = [\Psi_\mu, \Psi_\sigma]\) from \(h_s\); Pass 2 predicts non-anchors using both \(\Psi\) and the context feature \(\Phi = [\Phi_\mu, \Phi_\sigma]\) from \(g_{cm}\).

In the Slimmable Model, the active channels form a contiguous prefix of length \(w_y\) (the number of active latent channels). The hyperprior branch produces \(2w_y\) channels \((\Psi_\mu, \Psi_\sigma)\), and the context branch produces \(2w_y\) channels \((\Phi_\mu, \Phi_\sigma)\). To feed \(g_{ep}\) with correctly paired statistics, we replace Concat with Shuffling operator (Fig.~\ref{fig:shuffle}(bottom)), which interleaves the two branches as \((\Phi_\mu, \Phi_\sigma, \Psi_\mu, \Psi_\sigma)\) per active channel, yielding
\begin{equation}
\mathrm{Shuffle}(\Psi', \Phi') \in \mathbb{R}^{4w_y \times H_y \times W_y},
\label{eq:shuffle}
\end{equation}
with unused channels zero-padded. For \(g_{ep}\)'s first sCONV, the input index set is
\begin{equation}
\mathcal{I}_{\mathrm{in}}^{ep} = \{0, 1, \ldots, 4w_y - 1\},
\label{eq:index_ep}
\end{equation}
and the output index set is \(\mathcal{I}_{\mathrm{out}}^{ep} = \{0, 1, \ldots, 2w_y - 1\}\), ensuring paired \((\mu, \sigma)\) placement.

In the Mask Model, active channels are non-contiguous and given by \(\mathcal{I}_y \subseteq \{0, \ldots, C_y - 1\}\), with \(w_y = |\mathcal{I}_y|\). Sliced operators select channels from \(h_a\), \(h_s\), \(g_{cm}\), and \(g_{ep}\) using \(\mathcal{I}_y\), while index propagation (Sec.~\ref{sec:sliced}) preserves the \(\hat{y}\)-\((\mu, \sigma)\) pairings. For each \(i \in \mathcal{I}_y\), the four entropy-related channels fed into the first sCONV of \(g_{ep}\) are indexed by
\[
\mathcal{I}_{\mathrm{in}}^{ep} = \bigcup_{i\in \mathcal{I}_y}\{4i, 4i+1, 4i+2, 4i+3\},
\]
generalizing Eq.~\eqref{eq:index_ep} to non-contiguous channels. Shuffling is thus implicit via index propagation and tensor rearrangement into the interleaved order \((\Phi_\mu, \Phi_\sigma, \Psi_\mu, \Psi_\sigma)\), instead of an explicit operator. After M2S conversion, \(\mathcal{I}_y\) becomes a contiguous prefix of length \(w_y\), and the explicit shuffling operator above applies to the converted Slimmable Model.

\subsection{M2S Conversion}
\label{sec:reordering}

\begin{figure*}[t!]
\centering
  \setlength{\abovecaptionskip}{0.1cm}
  \setlength{\belowcaptionskip}{-0.4cm}
  \includegraphics[width=0.98\textwidth]{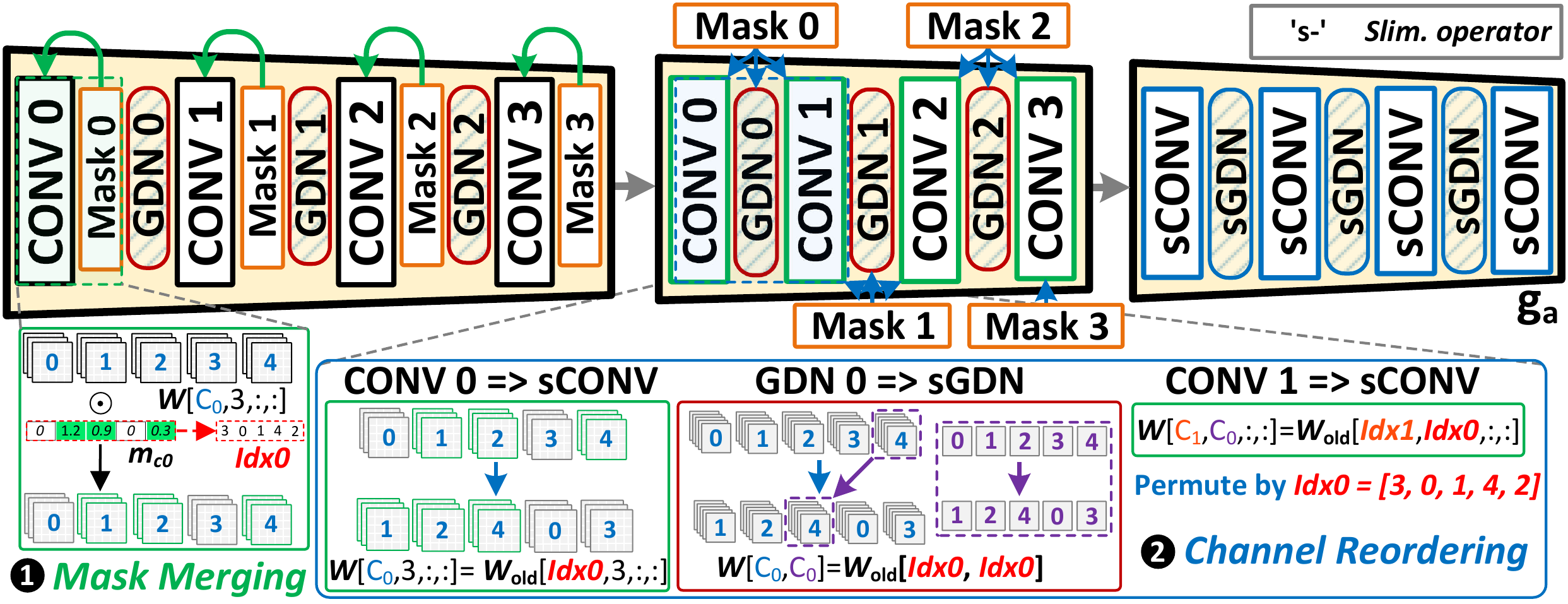}
\caption{M2S conversion from the trained Mask model (top-left) to the Slimmable model (top-right) via \ding{182} mask merging and \ding{183} channel reordering. Mask merging absorbs the scaling factors of Mask~\(l\) into the preceding CONV~\(l\) (lower-left/middle). Channel reordering permutes the output channels of CONV~\(l\), the input/output channels of GDN~\(l\), and the input channels of the next CONV~\(l+1\) using the index set \(\mathrm{Idx}_l\) derived from Mask~\(l\) (lower-right). The middle panel shows the intermediate state; \(\mathrm{Idx}_0/\mathrm{Idx}_1\) are derived from Mask~0/1, and \(C_0/C_1\) denote the output channel counts of the first two CONVs.}
\label{fig:reordering}
\end{figure*}

After training, \textbf{M2S} converts the Mask model into Slimmable model via \ding{182} mask merging and \ding{183} channel reordering (Fig.~\ref{fig:reordering}), removing the mask layer and making active channels contiguous for efficient inference.

\noindent\textbf{Mask merging.}
Each mask layer follows a CONV (or TPCV) and scales its output with \(s_c = \max(m_c, 0)\), as defined in Eq.~\ref{eq:mask_forward}. Since M2S conversion operates at the maximum width, all channels are retained. The scaling is absorbed into the preceding CONV as:
\begin{equation}
\mathbf{W}'_{i,j} = \mathbf{W}_{i,j} \cdot s_i,\quad b'_i = b_i \cdot s_i,
\label{eq:mask_merging}
\end{equation}
so the mask layer can be removed without changing the forward pass (Fig.~\ref{fig:reordering}, lower-left/middle).

\noindent\textbf{Channel reordering.}
Since the Mask model selects non-contiguous channels, we reorder channels by importance scores to make selected channels contiguous (Fig.~\ref{fig:reordering}). Each Mask~\(l\) yields an index set \(\mathrm{Idx}_l\), which determines the permutation of (i) the output channels of the preceding CONV~\(l\), (ii) the input and output channels of the following GDN~\(l\), and (iii) the input channels of the next CONV~\(l+1\). This per-layer strategy yields a transformation exact up to floating-point rounding errors introduced by merging.

M2S conversion yields a slimmable network that reuses existing tooling, turning non-contiguous Mask training into contiguous-prefix inference without dynamic indexing.

\subsection{Variable-Rate Modulation and Training}
\label{sec:modulation}

\noindent\textbf{Baseline VBR Framework:}
To enable continuous rate control, we integrate a modulation network that generates $\lambda$-conditioned scaling factors, following MAE approach \cite{yang2020variable}. The analysis and synthesis branches are lightweight MLPs:
\begin{equation}
\mathbf{g}_{\mathrm{enc}} = \exp(\mathrm{MLP}_{\mathrm{enc}}(\lambda)),\quad
\mathbf{g}_{\mathrm{dec}} = \exp(\mathrm{MLP}_{\mathrm{dec}}(\lambda)).
\label{eq:lambda_mod}
\end{equation}

These factors modulate the feature maps of \(g_a\) and \(g_s\) channel-wise, and the same network generates the quantization step sizes \(\boldsymbol{\Delta}_y\) and \(\boldsymbol{\Delta}_z\) for the Q variant below.

To avoid training instability caused by simultaneous width and rate adaptation, we adopt a progressive multi-stage training strategy.
In Stage~0, we train a Mask model (Sec.~\ref{sec:ops}) with a fixed $\lambda = \lambda_{\mathrm{max}}$ to learn non-contiguous channel importance. 
Stage~1 performs mask merging and channel reordering to convert the Mask model into a Slimmable model (Sec.~\ref{sec:reordering}), followed by fixed-$\lambda$ fine-tuning to obtain M2S model. 
In Stage~2, we fine-tune the converted model with random $\lambda$ sampling. The training objective is:
\begin{equation}
\underset{\lambda \sim \mathrm{Unif}(\Lambda)}{\mathbb{E}}
\sum_{w \in \mathcal{W}} \left( R_{w,\lambda} + \lambda D_{w,\lambda} \right),
\label{eq:train_obj}
\end{equation}
where \(\mathcal{W} = \{48,72,96,144,192\}\) is a predefined width set, and \(\mathrm{Unif}(\Lambda)\) denotes uniform sampling over \(\Lambda\). For each sampled \(\lambda\), the loss is summed over all widths in \(\mathcal{W}\). We refer to this baseline variable-rate configuration as \textbf{VBR}.

\noindent\textbf{Quantization-Step Scaling (Q):}
To further enhance rate adaptability, we introduce quantization-step scaling on the latent $y$ (or $z$ for Hyperprior), following QVRF~\cite{tong2023qvrf}:
\begin{equation}
\hat{y} = \mathrm{round}(y / \boldsymbol{\Delta}_y) \odot \boldsymbol{\Delta}_y.
\label{eq:q_step}
\end{equation}
where $\boldsymbol{\Delta}_y$ is the $\lambda$- and width-conditioned quantization step size, generated by the modulation network and broadcast to the channel dimension of $y$.
The scaled likelihood depends on the entropy model. For the factorized entropy model (used for $y$ in E2E and for $z$ in Hyperprior/CKBD),
\begin{equation}
p(\hat{y}_c \mid \boldsymbol{\Delta}_y) =
\int_{\hat{y}_c - \Delta_{y,c}/2}^{\hat{y}_c + \Delta_{y,c}/2}
p_{y_c}(\xi)\,\mathrm{d}\xi .
\label{eq:factorized_q}
\end{equation}
For the Gaussian conditional entropy model (used for $y$ in Hyperprior/CKBD), scaling $\hat{y}$ by $\boldsymbol{\Delta}_y$ scales both the predicted mean and scale. For each channel $c$,
\begin{equation}
p(\hat{y}_c \mid \hat{z}, \boldsymbol{\Delta}_y) =
\mathcal{N}\!\left(\mu_c \Delta_{y,c},\, \sigma_c \Delta_{y,c}\right)
* \mathrm{Unif}\!\left(-\frac{\Delta_{y,c}}{2},\, \frac{\Delta_{y,c}}{2}\right).
\label{eq:gaussian_q}
\end{equation}
and the full distribution factorizes over channels as
\(
p(\hat{y} \mid \hat{z}, \boldsymbol{\Delta}_y) = \prod_c p(\hat{y}_c \mid \hat{z}, \Delta_{y,c}).
\)
We denote this variant \textbf{VBR-Q}, sharing VBR's training pipeline. For CKBD, Eq.~\eqref{eq:gaussian_q} applies separately to anchors and non-anchors with independent step vectors. For numerical stability, the factorized model clamps its quantization step to \(10^{-2}\) during CDF construction to bound PMF bins, and the Gaussian conditional model lower-bounds its scale by the minimum scale-table entry; both affect only entropy-coding CDF tables, not forward quantization.

\begin{algorithm}[t!]
\caption{Training pipeline for VBR variants (illustrated with VBR-QFW). Each stage stops early when the learning rate decays to \(1.22\times10^{-8}\). \(\star\) lines mark the freezing/unfreezing operations.}
\label{alg:training}
\begin{algorithmic}[1]
\REQUIRE Model $\Theta$; width set $\mathcal{W}$; $\lambda$ set $\Lambda$; \\ maximum epochs \(E_{\max}\); warm-up epochs $E_w$.
\STATE Initialize $\Theta$ as \underline{Mask model} without modulation;
\STATE \textbf{Stage 0: Mask pretraining}
\FOR{$epoch = 1$ \TO $E_{\max}$}
    \STATE Fix $\lambda = \lambda_{\max}$; Sample batch $x$;
    \STATE $\mathcal{L} \gets \sum_{w \in \mathcal{W}} (R_{w,\lambda} + \lambda D_{w,\lambda})$;
    \STATE Back-propagate and update all parameters;
\ENDFOR
\STATE \textbf{Stage 1: Mask-to-Slimmable (M2S) conversion}
\STATE Perform mask merging and channel reordering;
\STATE Convert $\Theta$ into \underline{Slimmable model} without modulation;
\STATE Reset optimizer states; Fine-tune $\Theta$ for \(E_{\max}\) epochs;
\STATE \textbf{Stage 2: $\lambda$-based variable-rate fine-tuning}
\STATE Add $\lambda$-modulation network; Reset optimizer states;
\STATE Freeze \(g_a\), \(g_s\); \hfill \(\star\)
\FOR{$epoch = 1$ \TO \(E_{\max}\)}
    \STATE Sample \(\lambda \sim \mathrm{Unif}(\Lambda)\) and \(x\) from dataset;
    \STATE Compute \(\mathcal{L}\) with Eq.~\eqref{eq:train_obj};
    \STATE \textbf{if} $epoch = E_w+1$ \textbf{then} Unfreeze all parameters;\hfill \(\star\)
    \STATE Back-propagate and update all trainable parameters;
\ENDFOR
\STATE \textbf{Stage 3: \(w\)-based variable-rate fine-tuning}
\STATE Add $w$-modulation network; Reset optimizer states;
\STATE Freeze \(g_a\), \(g_s\), and \(\lambda\)-modulation network; \hfill \(\star\)
\STATE Execute the \textbf{Stage 2} loop for \(E_{\max}\) epochs;
\ENSURE \underline{Slimmable model} $\Theta$ with $\lambda/w$ modulation.
\end{algorithmic}
\end{algorithm}

\noindent\textbf{Freeze-Then-Fine-Tune Strategy (F):}
To prevent a randomly initialized modulation network from disrupting the width-adaptive structure learned by the Mask model, we freeze \(g_a\), \(g_s\), and the pre-trained \(\lambda\)-modulation network during warm-up, training only the entropy-related modules (\(h_a/h_s/g_{cm}/g_{ep}\)), then fine-tune the full network with random \(\lambda\) sampling. This variant uses the suffix \textbf{F} (e.g., \textbf{VBR-QF}).

\noindent\textbf{Width-Adaptive Modulation (W):}
The optimal feature modulation is coupled with the active width $w$, as a narrow model at a high bitrate requires different scaling than a wide model. We introduce a width-adaptive modulation network:
\begin{equation}
\mathbf{r}_{\mathrm{enc/dec}} = \exp(\mathrm{MLP}_{\mathrm{enc/dec}}(w_{\mathrm{norm}})).
\label{eq:width_mod}
\end{equation}
where \(w_{\mathrm{norm}} = (w-w_{\min})/(w_{\max}-w_{\min})\); the final modulation is the Hadamard product of these factors:
\begin{equation}
\mathbf{g}_{\mathrm{enc}}^{\mathrm{final}} = \mathbf{g}_{\mathrm{enc}} \odot \mathbf{r}_{\mathrm{enc}},\quad
\mathbf{g}_{\mathrm{dec}}^{\mathrm{final}} = \mathbf{g}_{\mathrm{dec}} \odot \mathbf{r}_{\mathrm{dec}}.
\label{eq:final_mod}
\end{equation}
Applied to VBR-QF, this yields \textbf{VBR-QFW}, where the warm-up freezes $g_a/g_s$ and the $\lambda$-modulation network. This variant is detailed in Algorithm~\ref{alg:training}.

\noindent\textbf{Width-Adaptive Modulation without Freezing (VBR-QW):}
Because F harms the tightly coupled CKBD context model, \textbf{VBR-QW} adds width-adaptive modulation to a pre-trained VBR-Q model but skips freezing, jointly fine-tuning all parameters with random \(\lambda\) and width sampling. It follows Algorithm~\ref{alg:training} without the \(\star\)-marked freeze/unfreeze steps and achieves the smallest RD gap to M2S on E2E and CKBD.

Table~\ref{tab:vbr_variants} summarizes the five variants. All share Stage~0 (Mask training) and Stage~1 (M2S conversion and fixed-\(\lambda\) fine-tuning). Stage~2 adds \(\lambda\)- modulation, and Stage~3 adds $w$-modulation. VBR-QW initializes from VBR-Q, adds $w$-modulation, and skips freezing.

\begin{table}[t]
\centering
\renewcommand{\arraystretch}{1.15}
\caption{Comparison of variable-rate training variants. 
}
\label{tab:vbr_variants}
\begin{tabular}{l|ccccc}
\hline \hline
Variant & \(\lambda\)-mod. & Q-step scaling & \(w\)-mod. & Freeze & Stage \\
\hline
VBR     & \checkmark &            &            &            & 2 \\
VBR-Q   & \checkmark & \checkmark &            &            & 2 \\
VBR-QF  & \checkmark & \checkmark &            & \checkmark & 2 \\
VBR-QW  & \checkmark & \checkmark & \checkmark &            & 3 \\
VBR-QFW & \checkmark & \checkmark & \checkmark & \checkmark & 3 \\
\hline \hline
\end{tabular}
\end{table}

%% file: sections/eval.tex
\section{EVALUATION}
\label{sec:eval}

\begin{figure*}[t!]
\centering
  \setlength{\abovecaptionskip}{0.1cm}
  \setlength{\belowcaptionskip}{-0.4cm}
\subfloat{\includegraphics[width=0.99\textwidth]{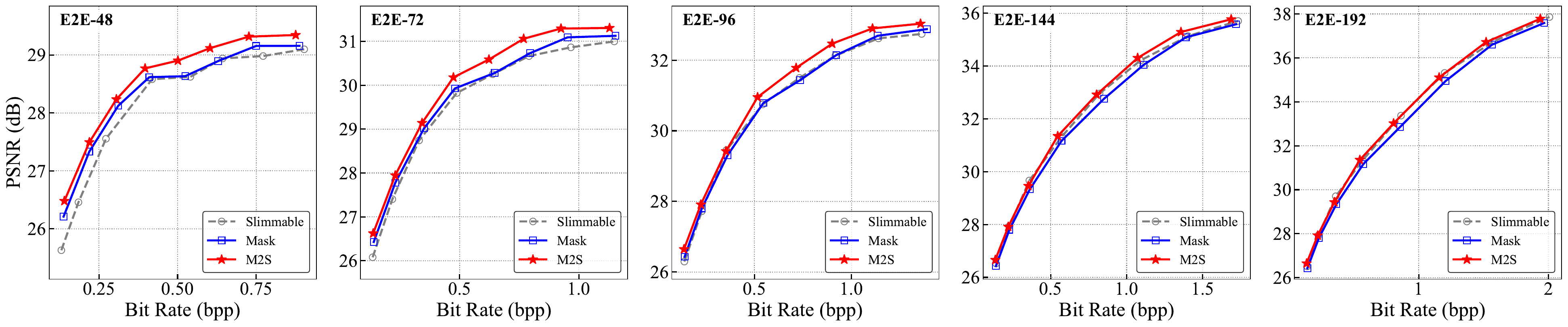}
\label{fig:rd_comparison_e2e}}
\vspace{-0.3cm}
\subfloat{\includegraphics[width=0.99\textwidth]{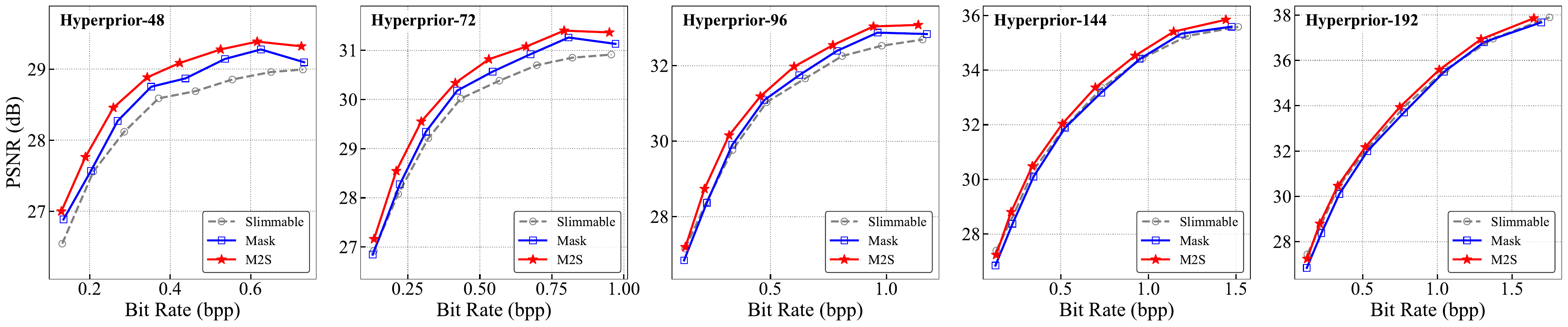}
\label{fig:rd_comparison_hyper}}
\vspace{-0.3cm}
\subfloat{\includegraphics[width=0.99\textwidth]{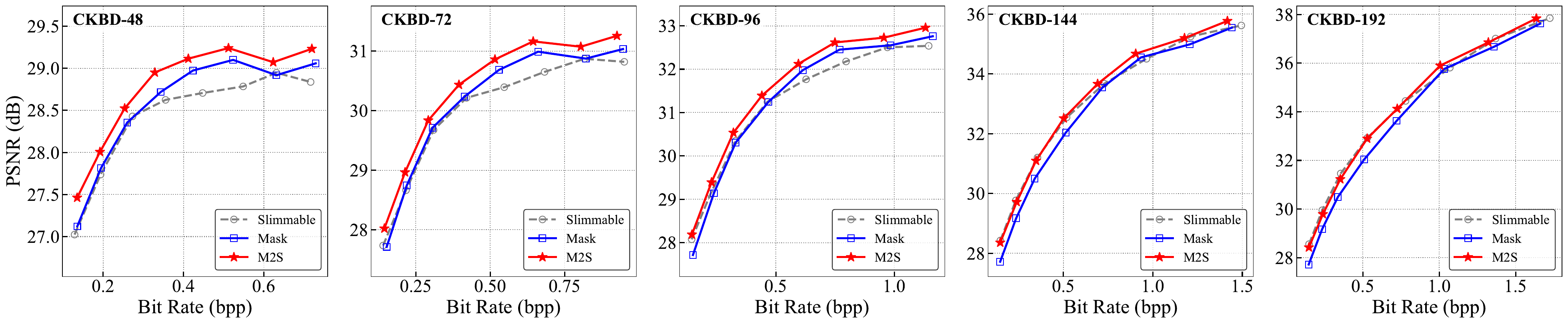}
\label{fig:rd_comparison_ckbd}}
\caption{RD performance comparison of Slimmable, Mask, and M2S on Kodak across five widths (48--192) for E2E, Hyperprior and CKBD backbones. M2S consistently outperforms both baselines at narrow widths (48, 72), while eliminating the regularization overhead of Mask at wide width (144, 192).}
\label{fig:rd_comparison}
\end{figure*}

\subsection{Experimental Setup}
\label{sec:exp_setup}

\noindent\textbf{Datasets.} We train on CLIC 2020 and DIV2K HR\footnote{https://data.vision.ee.ethz.ch/cvl/DIV2K/} (2357 images), randomly cropping $256\times256$ patches with batch size 8. For the models compared in Section~\ref{sec:eval_others}, we additionally include the Flickr 2W dataset (20745 images) during the width-based modulation fine-tuning stage. CLIC 2021 professional validation\footnote{https://archive.compression.cc/2021/tasks/index.html} is used for validation, and Kodak (24 images, $768\times512$) for evaluation. Additional test sets (e.g., Tecnick) are not included due to space constraints; results can be added in a revision if requested.

\noindent\textbf{Compared Methods.} We compare with (1) traditional codecs: JPEG, JPEG2000, BPG, and VVC (VTM 19.2) intra-mode; and (2) elastic LIC methods: CBANet \cite{guo2023cbanet}, EVC \cite{guoevc}, and ELFIC \cite{zhang2023elfic}, using publicly released pretrained models. All models are evaluated in PyTorch 2.6.0 with CompressAI 1.2.8.

\noindent\textbf{Implementation Details.} The model is trained and evaluated on this predefined width set $\mathcal{W}=\{48,72,96,144,192\}$; extending to other widths would require including them in \(\mathcal{W}\) during training or fine-tuning, without architectural changes. We use Adam with initial learning rates $10^{-4}$ (main network) and $10^{-6}$ (auxiliary entropy parameters), and $\Lambda=\{18,35,67,130,250,483,932,1800\}\times10^{-4}$. Training runs for 800 epochs until convergence, using \texttt{ReduceLROnPlateau} (factor $0.5$, patience $10$); convergence is reached when the learning rate decays to $1.22\times10^{-8}$. Single-$\lambda$ models use a fixed $\lambda\in\Lambda$. 
Variable-rate models adopt the progressive training strategy in Section~\ref{sec:modulation}. Freeze-then-fine-tune variants (e.g., VBR-QF) use a 100-epoch warm-up in Stage 2 ($\lambda$-modulation), and VBR-QFW additionally uses a 100-epoch warm-up in Stage 3 ($w$-modulation). For the models compared in Section~\ref{sec:eval_others}, the Stage 3 warm-up is reduced to 5 epochs.

\noindent\textbf{Hardware.} Training: Intel Xeon Platinum 8457C + Nvidia A100 80GiB. Evaluation: Intel Xeon 48-core E5-2650 + Nvidia RTX3080 10GiB.

\noindent\textbf{Metrics.} We report bitrate (bpp), PSNR (dB), and MS-SSIM\footnote{MS-SSIM (dB) $= -10\log_{10}(1-\text{MS-SSIM})$} \cite{wang2003multiscale}. RD curves compare methods, with BD-rate \cite{bjontegaard2001calculation} as average bitrate saving at equivalent quality.

\subsection{Slimmable vs Mask vs M2S}
\label{sec:comparison}

We compare three variants: \textbf{Slimmable} model, \textbf{Mask} model, and \textbf{M2S} model (in Sec.~\ref{sec:reordering}).
All variants follow the training protocol in Sec.~\ref{sec:exp_setup} and share the same width set $\mathcal{W}$. We evaluate E2E, Hyperprior, and CKBD backbones to demonstrate the generality of our approach.

\noindent\textbf{Narrow widths ($w=48, 72$).} For all three backbones, Mask outperforms Slimmable at narrow widths, especially at low bitrates, confirming that non-contiguous selection preserves the most informative channels. M2S surpasses both, achieving the highest PSNR at every bitrate point. At $w=48$, M2S achieves BD-rate savings over Slimmable of $19.22\%$, $24.11\%$, and $22.76\%$ for E2E, Hyperprior, and CKBD, respectively; over Mask, the savings are $10.54\%$, $16.24\%$, and $17.08\%$. Fig.~\ref{fig:rd_comparison} shows the RD curves, where M2S dominates both baselines across all five widths for E2E and Hyperprior, and across narrow and intermediate widths for CKBD.

\begin{table}[t!]
\centering
\renewcommand{\arraystretch}{1.1}
\caption{BD-rate savings (\%) compared to the same-width Slimmable baseline on Kodak.}
\begin{tabular}{c|c|ccc}
\hline \hline
 & Width & Mask (vs Slim) & M2S (vs Slim) & M2S (vs Mask) \\
\hline
\multirow{5}{*}{\textbf{E2E}} & 48 & -10.60 & -19.22 & -10.54 \\
                      & 72 & -5.73 & -14.44 & -9.67 \\
                      & 96 & +0.26 & -8.33 & -8.93 \\
                      & 144 & +5.71 & -2.77 & -8.01 \\
                      & 192 & +6.73 & -1.29 & -7.54 \\ \hline
\multirow{5}{*}{\textbf{Hyper}} & 48 & -11.23 & -24.11 & -16.24 \\
                      & 72 & -7.33 & -18.98 & -12.82 \\
                      & 96 & -2.36 & -13.52 & -11.91 \\
                      & 144 & +3.72 & -6.37 & -9.79 \\
                      & 192 & +5.95 & -3.59 & -9.08 \\ \hline
\multirow{5}{*}{\textbf{CKBD}} &  48 & -5.81 & -22.76 & -17.08 \\
                      & 72 & -4.63 & -17.58 & -12.68 \\
                      & 96 & +0.36 & -11.25 & -11.94 \\
                      & 144 & +11.64 & -1.59 & -11.87 \\
                      & 192 & +14.11 & +1.06 & -11.48 \\
\hline \hline
\end{tabular}
\label{tab:bdrate}
\end{table}

\noindent\textbf{Intermediate widths ($w=96, 144$).} At $w=96$, Mask and Slimmable perform similarly for E2E and Hyperprior, while CKBD Mask is already slightly worse ($+0.36\%$). At $w=144$, Mask degrades for all three backbones due to regularization overhead, with CKBD most severe ($+11.64\%$), yet M2S still recovers to $-1.59\%$. Hyperprior exhibits slightly larger gains than E2E due to stronger entropy modeling, while CKBD benefits most from M2S conversion.

\noindent\textbf{Full width ($w=192$).} Mask underperforms Slimmable at high bitrates (e.g., $\sim$0.27 dB at 2.0 bpp for E2E) due to channel-wise regularization. M2S eliminates this for E2E and Hyperprior, recovering to $-1.29\%$ and $-3.59\%$ BD-rate savings, respectively. For CKBD, M2S reduces the degradation from $+14.11\%$ to $+1.06\%$, i.e., still slightly worse than Slimmable but far better than Mask ($-11.48\%$ vs Mask). Overall, M2S largely mitigates the regularization overhead across all three backbones.

\begin{figure*}[t]
  \setlength{\abovecaptionskip}{0.1cm}
  \setlength{\belowcaptionskip}{-0.4cm}
\centering
\includegraphics[width=1.01\textwidth]{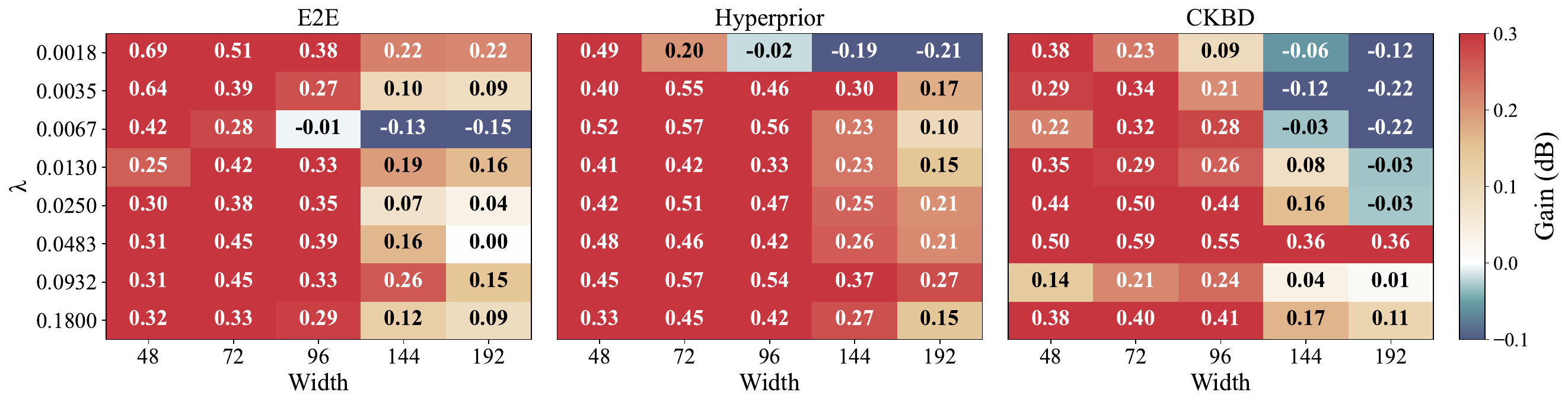}
\caption{PSNR gain (dB) of M2S over Slimmable across all widths and $\lambda$ values on Kodak. Each cell is obtained by cubic-spline interpolation of the RD curves at the minimum bpp for the given \(\lambda\) and width.}
\label{fig:heatmap}
\end{figure*}

\noindent\textbf{Overall.} Fig.~\ref{fig:heatmap} shows M2S gains over Slimmable. E2E peaks at $0.69$~dB ($w=48$, $\lambda=0.0018$), Hyperprior shows similar but larger gains (consistent with Table~\ref{tab:bdrate}), and CKBD peaks at $0.59$~dB ($w=72$, $\lambda=0.0483$), all concentrated at narrow widths and low bitrates. Gains diminish as width increases and approach zero or negative at $w=144,192$, confirming that non-contiguous selection benefits slimmable training most when the channel budget is tight, and that M2S preserves narrow-width gains while eliminating regularization overhead.

\subsection{Variable-Rate Performance}
\label{sec:var_rate_eval}

\begin{figure*}[t]
\centering
\subfloat{\includegraphics[width=0.99\textwidth]{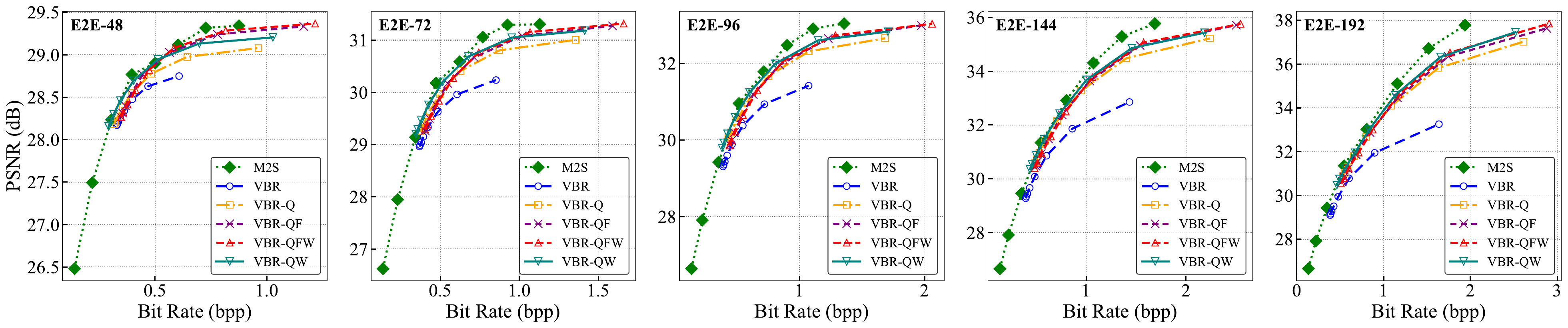}}
\vspace{-0.4cm}
\subfloat{\includegraphics[width=0.99\textwidth]{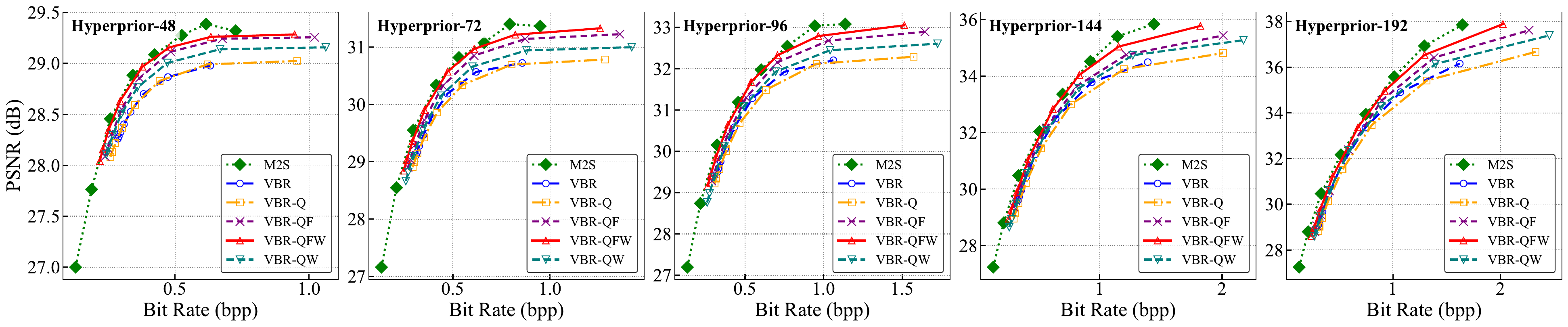}}
\vspace{-0.4cm}
\subfloat{\includegraphics[width=0.99\textwidth]{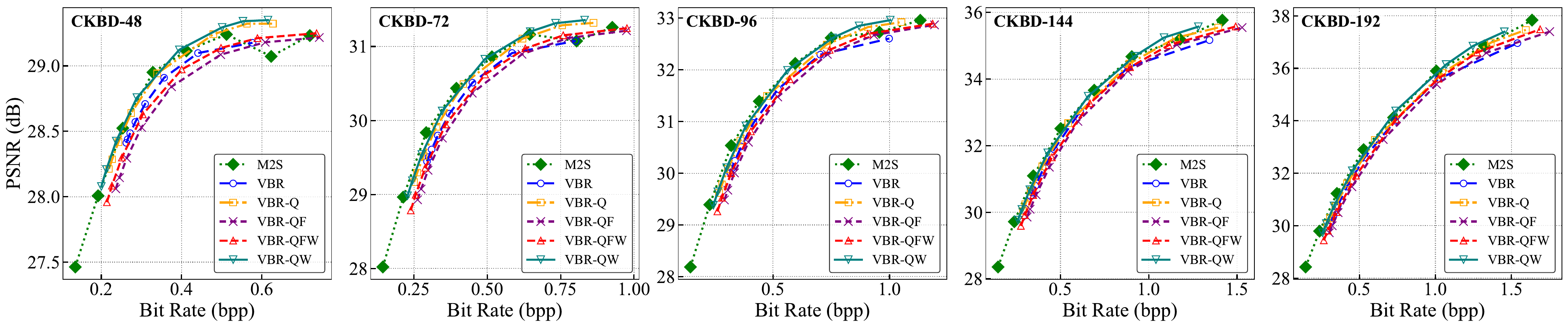}}
\caption{RD performance comparison across five widths (48--192) on Kodak for (a) E2E, (b) Hyperprior, and (c) CKBD backbones. VBR-QW achieves the best performance on E2E and CKBD, while VBR-QFW remains the best on Hyperprior.}
\label{fig:rd_vbr_comparison}
\end{figure*}

We compare five variable-rate variants: VBR, VBR-Q, VBR-QW, VBR-QF, and VBR-QFW, as defined in Sec.~\ref{sec:modulation}. The fixed-rate M2S is included as an upper bound. For E2E backbone and Hyperprior backbone, the variable-rate evaluations of parameter and latency are provided in the supplementary material.

\subsubsection{Rate-Distortion Performance}

\begin{table}[t!]
\centering
\renewcommand{\arraystretch}{1.15}
\caption{BD-rate increase (\%) over the M2S upper bound on Kodak. Best results per backbone are in bold.}
\label{tab:bdrate_vbr_compare}
\begin{tabular}{c|c|ccccc}
\hline \hline
& Width & 48 & 72 & 96 & 144 & 192 \\
\hline
\multirow{5}{*}{\textbf{E2E}} & VBR     & 21.57 & 27.64 & 28.36 & 31.28 & 36.92 \\
                              & VBR-Q   & 15.86 & 16.58 & 14.75 & 14.18 & 16.47 \\
                              & VBR-QW  & \textbf{4.85} & \textbf{7.30} & \textbf{6.92} & \textbf{7.82} & \textbf{9.68} \\
                              & VBR-QF  & 10.79 & 17.61 & 16.51 & 14.54 & 16.17 \\
                              & VBR-QFW & 10.37 & 17.36 & 16.70 & 14.53 & 14.84 \\ \hline
\multirow{5}{*}{\textbf{Hyper}} & VBR     & 30.25 & 24.79 & 20.43 & 18.59 & 19.00 \\
                               & VBR-Q   & 29.34 & 31.09 & 27.42 & 26.99 & 28.80 \\
                               & VBR-QW  & 16.78 & 19.75 & 18.30 & 19.12 & 20.09 \\
                               & VBR-QF  & 11.32 & 12.29 & 13.10 & 15.08 & 15.30 \\
                               & VBR-QFW & \textbf{4.29} & \textbf{4.12} & \textbf{5.50} & \textbf{6.83} & \textbf{7.88} \\ \hline
\multirow{5}{*}{\textbf{CKBD}} & VBR     & 5.55 & 10.48 & 11.78 & 8.43 & 9.04 \\
                               & VBR-Q   & -4.33 & 2.36 & 4.62 & 4.24 & 4.73 \\
                               & VBR-QW  & \textbf{-6.17} & \textbf{0.01} & \textbf{1.90} & \textbf{1.35} & \textbf{1.32} \\
                               & VBR-QF  & 16.48 & 18.69 & 17.14 & 14.08 & 14.50 \\
                               & VBR-QFW & 9.60 & 13.88 & 13.07 & 10.76 & 10.73 \\
\hline \hline
\end{tabular}
\end{table}

Fig.~\ref{fig:rd_vbr_comparison} presents the RD curves across five widths for the three backbones, and Table~\ref{tab:bdrate_vbr_compare} reports the BD-rate increase over the fixed-rate M2S upper bound. The effectiveness of the variable-rate components is backbone-dependent.

\noindent\textbf{E2E backbone.} VBR shows the largest gap to M2S (21.6--36.9\%). VBR-Q reduces it to 14.2--16.6\% via quantization-step scaling, and VBR-QF further reduces it to 10.8--17.6\% via freeze-then-fine-tune. VBR-QW achieves the best performance, narrowing the gap to 4.9--9.7\%. VBR-QFW is slightly worse than VBR-QW, indicating that width-modulation on top of freeze-then-fine-tune does not bring further gains on E2E.

\noindent\textbf{Hyperprior backbone.} VBR-Q is slightly better than VBR only at the narrowest width ($29.34\%$ vs. $30.25\%$ at $w=48$), but worse at all wider widths. VBR-QF brings significant gains, reducing the gap to $11.3$--$15.3\%$, confirming that the freeze-then-fine-tune strategy is effective when the entropy model is loosely coupled with the transform. VBR-QFW achieves the best performance on Hyperprior, reducing the gap to $4.1$--$7.9\%$.

\noindent\textbf{CKBD backbone.} VBR-QF is the worst variant (14.1--18.7\%), even worse than VBR (5.6--11.8\%), because freezing $g_a/g_s$ misaligns the tightly coupled checkerboard context model. VBR-Q avoids this by skipping freezing, achieving $-4.3\%$ to $4.7\%$. VBR-QW is the best, with the gap ranging from $-6.2\%$ to $1.9\%$.

\noindent\textbf{Overall.} VBR-QW is the most consistently effective variant, achieving the best results on E2E and CKBD. On Hyperprior, VBR-QFW is best, followed by VBR-QF, while VBR-QW ranks third. Freeze-then-fine-tune brings clear gains on Hyperprior and modest gains on E2E, but harms CKBD, highlighting the need for backbone-aware design. At low bitrates ($\lambda \leq 0.0067$), all variants perform similarly as the rate constraint dominates.

\begin{table*}[t]
\centering
\renewcommand{\arraystretch}{1.15}
\caption{Parameter comparison of the CKBD backbone across different
widths. The CKBD row reports single-$\lambda$
single-width models; the M2S row reports one model covering all
five widths at a fixed $\lambda$; the VBR variants cover 5 widths
$\times$ 8 $\lambda$ values with a single model.
``Total (8 $\lambda$s)'' reports the total storage required if all
eight $\lambda$ values were deployed, i.e., $8\times$ the per-model
total for single-$\lambda$ models, and a single model's total for the
VBR variants.}
\label{tab:vbr_params}
\begin{tabular}{c|c|cccccccccc}
\hline \hline
    &      & \multicolumn{10}{c}{\textbf{Params}}      \\ \cline{3-12} 
\multirow{-2}{*}{\textbf{Model}} & \multirow{-2}{*}{\textbf{Width}} & \multicolumn{1}{c|}{\textbf{$g_a$}}            & \multicolumn{1}{c|}{\textbf{$g_s$}}            & \multicolumn{1}{c|}{\textbf{$h_a$}}            & \multicolumn{1}{c|}{\textbf{$h_s$}}            & \multicolumn{1}{c|}{\textbf{FEM}} & \multicolumn{1}{c|}{\textbf{$g_{cm}$}} & \multicolumn{1}{c|}{\textbf{$g_{ep}$}}           & \multicolumn{1}{c|}{\textbf{ModNet}}     & \multicolumn{1}{c|}{\textbf{Total}} & \textbf{Total (8 $\lambda$s)} \\ \hline
                                 & 48                               & \multicolumn{1}{c|}{183.65 K}                 & \multicolumn{1}{c|}{183.60 K}                 & \multicolumn{1}{c|}{136.08 K}                 & \multicolumn{1}{c|}{255.98 K}                 & \multicolumn{1}{c|}{2.93 K}                              & \multicolumn{1}{c|}{115.30 K}                              & \multicolumn{1}{c|}{78.77 K}                  & \multicolumn{1}{c|}{}                    & \multicolumn{1}{c|}{956.36 K}       &                            \\ \cline{2-9} \cline{11-11}
                                 & 72                               & \multicolumn{1}{c|}{410.26 K}                 & \multicolumn{1}{c|}{410.19 K}                 & \multicolumn{1}{c|}{306.07 K}                 & \multicolumn{1}{c|}{575.78 K}                 & \multicolumn{1}{c|}{4.39 K}                              & \multicolumn{1}{c|}{259.34 K}                              & \multicolumn{1}{c|}{176.90 K}                 & \multicolumn{1}{c|}{}                    & \multicolumn{1}{c|}{2.14 M}         &                            \\ \cline{2-9} \cline{11-11}
                                 & 96                               & \multicolumn{1}{c|}{726.72 K}                 & \multicolumn{1}{c|}{726.63 K}                 & \multicolumn{1}{c|}{544.03 K}                 & \multicolumn{1}{c|}{1.02 M}                   & \multicolumn{1}{c|}{5.86 K}                              & \multicolumn{1}{c|}{460.99 K}                              & \multicolumn{1}{c|}{314.21 K}                 & \multicolumn{1}{c|}{}                    & \multicolumn{1}{c|}{3.80 M}         &                            \\ \cline{2-9} \cline{11-11}
                                 & 144                              & \multicolumn{1}{c|}{1.63 M}                   & \multicolumn{1}{c|}{1.63 M}                   & \multicolumn{1}{c|}{1.224 M}                  & \multicolumn{1}{c|}{2.30 M}                   & \multicolumn{1}{c|}{8.78 K}                              & \multicolumn{1}{c|}{1.037 M}                               & \multicolumn{1}{c|}{706.32 K}                 & \multicolumn{1}{c|}{}                    & \multicolumn{1}{c|}{8.54 M}         &                            \\ \cline{2-9} \cline{11-11}
\multirow{-5}{*}{\textbf{CKBD}}  & 192                              & \multicolumn{1}{c|}{2.89 M}                   & \multicolumn{1}{c|}{2.89 M}                   & \multicolumn{1}{c|}{2.18 M}                   & \multicolumn{1}{c|}{4.09 M}                   & \multicolumn{1}{c|}{11.71 K}                             & \multicolumn{1}{c|}{1.84 M}                                & \multicolumn{1}{c|}{1.26 M}                   & \multicolumn{1}{c|}{\multirow{-5}{*}{0}} & \multicolumn{1}{c|}{15.16 M}        & \multirow{-5}{*}{244.80 M} \\ \hline
\textbf{M2S}                     & 48-192                           & \multicolumn{1}{c|}{2.89 M}                   & \multicolumn{1}{c|}{2.89 M}                   & \multicolumn{1}{c|}{2.18 M}                   & \multicolumn{1}{c|}{4.09 M}                   & \multicolumn{1}{c|}{33.67 K}                             & \multicolumn{1}{c|}{1.84 M}                                & \multicolumn{1}{c|}{1.26 M}                   & \multicolumn{1}{c|}{0}                   & \multicolumn{1}{c|}{15.16 M}        & 121.29 M                   \\ \hline
\textbf{VBR}                     &                                  & \multicolumn{1}{c|}{}                         & \multicolumn{1}{c|}{}                         & \multicolumn{1}{c|}{}                         & \multicolumn{1}{c|}{}                         & \multicolumn{1}{c|}{}                                    & \multicolumn{1}{c|}{}                                      & \multicolumn{1}{c|}{}                         & \multicolumn{1}{c|}{224.64 K}            & \multicolumn{1}{c|}{15.41 M}        & 15.41 M                    \\ \cline{1-1} \cline{10-12} 
\textbf{VBR-Q}                   &                                  & \multicolumn{1}{c|}{}                         & \multicolumn{1}{c|}{}                         & \multicolumn{1}{c|}{}                         & \multicolumn{1}{c|}{}                         & \multicolumn{1}{c|}{}                                    & \multicolumn{1}{c|}{}                                      & \multicolumn{1}{c|}{}                         & \multicolumn{1}{c|}{336.96 K}            & \multicolumn{1}{c|}{15.52 M}        & 15.52 M                    \\ \cline{1-1} \cline{10-12} 
\textbf{VBR-QW}                  & \multirow{-3}{*}{48-192}         & \multicolumn{1}{c|}{\multirow{-3}{*}{2.89 M}} & \multicolumn{1}{c|}{\multirow{-3}{*}{2.89 M}} & \multicolumn{1}{c|}{\multirow{-3}{*}{2.18 M}} & \multicolumn{1}{c|}{\multirow{-3}{*}{4.09 M}} & \multicolumn{1}{c|}{\multirow{-3}{*}{33.67 K}}           & \multicolumn{1}{c|}{\multirow{-3}{*}{1.84 M}}              & \multicolumn{1}{c|}{\multirow{-3}{*}{1.26 M}} & \multicolumn{1}{c|}{673.92 K}            & \multicolumn{1}{c|}{15.84 M}        & 15.84 M                    \\ \hline \hline
\multicolumn{12}{l}{\footnotesize FEM: factorized entropy model; ModNet: modulation network.} \\
\multicolumn{12}{l}{\footnotesize VBR-Q/VBR-QF and VBR-QW/VBR-QFW share the same architecture within each pair, and thus identical complexity.} \\
\end{tabular}
\end{table*}

\begin{table*}[t]
\centering
\renewcommand{\arraystretch}{1.1}
\caption{Per-module latency (ms) of the CKBD (VBR-QW). ``Enc/Dec ent\_z'' and ``Enc/Dec ent\_y'' denote the encoding/decoding latency of the factorized entropy model for \(z\) and the Gaussian conditional entropy model for \(y\), respectively. The last column is the one-time CDF construction cost, paid only when \(\lambda/w\) changes.}
\label{tab:vbr_latency}
\begin{tabular}{c |c c c c c c |c c c c |c }
\hline \hline
\textbf{Width} & \textbf{\(g_a\)} & \textbf{\(h_a\)} & \textbf{\(h_s\)} & \textbf{\(g_s\)} & \textbf{\(g_{cm}\)} & \textbf{\(g_{ep}\)} & \textbf{Enc ent\_z} & \textbf{Enc ent\_y} & \textbf{Dec ent\_z} & \textbf{Dec ent\_y} & \textbf{CDF (startup)} \\ \hline
48 & 1.52 &	0.52 & 0.58 & 1.94 & 0.31 & 0.65 & 1.31 & 20.80 & 1.55 & 24.77 & 10.57--10.74 \\ \hline
72	& 2.28 & 0.54 & 0.59 & 2.89 & 0.31 & 0.68 & 1.79 & 23.65 & 2.18 & 29.50 & 14.66--15.59 \\ \hline
96	& 2.71 & 0.65 & 0.71 & 3.20 & 0.40 & 0.83 & 2.63 & 26.94 & 3.20 & 34.51 & 20.74--22.72 \\ \hline
144	& 5.36 	& 0.54 & 0.68 & 5.50 & 0.34 & 0.66 & 6.14 & 33.14 & 7.32 & 43.81 & 38.61--58.60 \\ \hline
192	& 6.84 	& 0.54 & 0.76 & 6.94 & 0.48 & 0.67 & 14.04 & 40.10 & 17.28 & 54.37 & 106.90--290.51 \\ 
\hline \hline
\end{tabular}
\end{table*}

\subsubsection{Parameter and Latency Analysis}

CKBD subsumes E2E/Hyperprior operators and has the largest parameter/entropy-coding footprint, so Tables~\ref{tab:vbr_params}--\ref{tab:vbr_latency} are upper bounds; per-backbone details are in supplementary material.

\noindent\textbf{Parameter efficiency.}
Table~\ref{tab:vbr_params} compares the parameter budget on CKBD, whose operator set covers E2E and Hyperprior. The backbone (\(g_a\), \(g_s\), \(h_a\), \(h_s\), FEM, \(g_{cm}\), \(g_{ep}\)) is shared across VBR variants, so only the modulation network differs. ModNet grows from 0 (M2S) to 224.64K (VBR), 336.96K (VBR-Q/QF), and 673.92K (VBR-QW/QFW), i.e., 1.5\%, 2.2\%, and 4.6\% overhead over the 15.16M M2S baseline. 
The 112.32K increase from VBR to VBR-Q on CKBD comprises three modulation layers: one for \(\Delta_y\) of anchors, one for \(\Delta_y\) of non-anchors, and one for \(\Delta_z\), whereas E2E requires only the \(\Delta_y\) layer.
One VBR-QW/QFW model (15.84M) replaces 40 independent models (244.80M), a 15.4$\times$ storage reduction; VBR and VBR-Q/QF achieve 15.9$\times$ and 15.8$\times$. E2E/Hyperprior follow the same trend with smaller backbones (5.816M/12.084M) and higher relative overhead (3.9/4.5/9.0\% and 1.9/2.2/4.3\%).

\noindent\textbf{Latency.}
Table~\ref{tab:vbr_latency} reports per-module latency on CKBD (VBR-QW), which covers all E2E/Hyperprior operators. At \(w=192\), transform networks cost only 16.2~ms, while entropy coding dominates: 54.1~ms encoding and 71.7~ms decoding for \(y\) and \(z\). Decoding is higher because CompressAI's \texttt{decompress} allocates a new GPU tensor per sample, causing CPU-to-GPU transfer. CDF construction is a one-time startup cost paid only when \(\lambda/w\) changes, growing from 10.6--10.7~ms at \(w=48\) to 106.9--290.5~ms at \(w=192\), driven by the \(\lambda\)-dependent \(z\) step and amplified by Flickr 2W fine-tuning. Modulation costs 1.2--4.6~ms. Both CDF construction and modulation are configuration-dependent and amortized in steady state, while GPU-side transforms remain stable.

\noindent\textbf{Overall.}
The modulation network adds only 1.5--4.6\% parameters on CKBD (similar on E2E/Hyperprior), while one model replaces 40 independent models, cutting storage by 14.8--15.9$\times$. Latency is dominated by entropy coding and CDF construction, which scale with width and amortize per image. Since entropy coding scales with active channels, channel slimming offers a practical latency--RD trade-off in one model.

\subsection{Comparison with Existing Elastic LIC Methods}
\label{sec:eval_others}

\begin{figure*}[t]
  \setlength{\abovecaptionskip}{0.1cm}
  \setlength{\belowcaptionskip}{-0.2cm}
\centering
\includegraphics[width=0.99\textwidth]{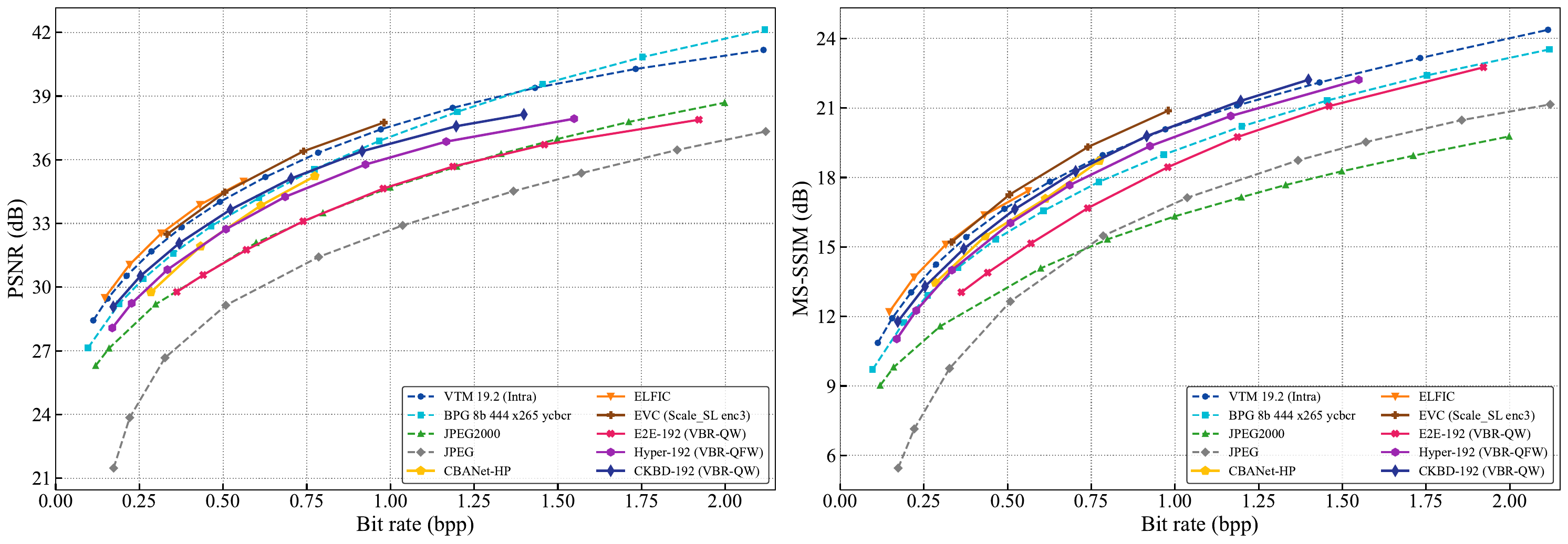}
\caption{Rate-distortion comparison on Kodak. E2E/Hyper/CKBD-192 denote variable-rate models at width 192. For CBANet, ELFIC, and EVC, the highest-quality RD curves among available models are used for reference.}
\label{fig:psnr_compare}
\end{figure*}

Following the taxonomy in Sec.~\ref{sec:intro}, the compared methods include slimmable (ELFIC~\cite{zhang2023elfic}) and scalable (EVC~\cite{guoevc}) coding, while CBANet~\cite{guo2023cbanet} is a variable-rate method that is neither. The ``One is all'' framework~\cite{li2025one} is excluded due to unavailable public code and pretrained models.
Fig.~\ref{fig:psnr_compare} shows the RD curves on Kodak, and Table~\ref{tab:others_latency} reports parameter count and en-/de-coding latency.

\noindent\textbf{Backbone configuration.}
Our CKBD backbone follows the Minnen2018-style checkerboard architecture \cite{minnen2018joint,he2021checkerboard} with a mean-scale Gaussian entropy model and plain convolutional transforms. Our CKBD-192 matches its baseline in Fig.~7 of \cite{he2021checkerboard}, validating our implementation. The identical backbone ensures a fair comparison, and the gap to Cheng2020-style models \cite{cheng2020learned} is orthogonal to our contributions.

\begin{table*}[t!]
\centering
\renewcommand{\arraystretch}{1.1}
\caption{Parameter count and en-/de-coding latency on Kodak.
\textit{Rel. Cplx.} (\%) is method-specific: \(w/192\times100\%\) (ours), the decoder-computation ratio (ELFIC, CBANet), and \(N_{\mathrm{enc}}/4\times100\%\) (EVC).
\(N_\lambda\) is the number of default variable-rate configurations; ``Interp.'': extra RD points via interpolation, not applicable to CBANet (independent bitrate-control modules).}
\label{tab:others_latency}
\begin{tabular}{c|ccc|c|c|c}
\hline \hline
    & \multicolumn{3}{c|}{\textbf{Ours}}   &     &     &     \\ \cline{2-4}
\multirow{-2}{*}{}    & \multicolumn{1}{c|}{\textbf{E2E}}   & \multicolumn{1}{c|}{\textbf{Hyper}} & \textbf{CKBD}  & \multirow{-2}{*}{\textbf{\begin{tabular}[c]{@{}c@{}}ELFIC\\ (Low Rate)\end{tabular}}} & \multirow{-2}{*}{\textbf{CBANet-HP}} & \multirow{-2}{*}{\textbf{\begin{tabular}[c]{@{}c@{}}EVC\\ (Scale\_SL)\end{tabular}}} \\ \hline
\textbf{Params (M)}                             & \multicolumn{1}{c|}{6.34}           & \multicolumn{1}{c|}{12.68}          & 15.84          & 66.99                                                                                 & 6.39                                 & 17.47                                                                                \\ \hline
\textbf{$N_{\lambda}$}                          & \multicolumn{3}{c|}{8 (Interp.)}                                                           & 4 (Interp.)                                                                           & 4                                    & 4 (Interp.)                                                                          \\ \hline
{\color[HTML]{0F1115} \textbf{Rel. Cplx. (\%)}} & \multicolumn{3}{c|}{25/37.5/50/75/100}                                                     & 7.66/35.82/74.06/100                                                                  & 25/50/100                            & 25/50/75/100                                                                         \\ \hline
\textbf{Enc. (ms)}                              & \multicolumn{1}{c|}{10/14/18/27/36} & \multicolumn{1}{c|}{19/24/28/40/51} & 28/32/37/51/71 & 166                                                                                   & 36                                   & 42/45/50/55                                                                          \\ \hline
\textbf{Dec. (ms)}                              & \multicolumn{1}{c|}{14/20/25/38/50} & \multicolumn{1}{c|}{23/31/36/51/66} & 32/38/45/62/86 & 175/193/199/202                                                                       & 54/56/58                             & 32                                                                                   \\ \hline \hline
\end{tabular}
\end{table*}

\noindent\textbf{Variable-rate capability (\(N_\lambda\)).}
Our single VBR model provides \(N_\lambda=8\) default configurations with interpolation, covering a wider bitrate range than all compared elastic LIC methods. For example, Hyper-192 spans 0.17--1.55\,bpp with 28.08--37.93\,dB PSNR and 12.5--24.5\,dB MS-SSIM. CBANet-HP has \(N_\lambda=4\) but cannot interpolate, because it switches independent BAL/IBAL modules on both sides of \(y\); its range is only 0.28--0.77\,bpp. ELFIC and EVC both have \(N_\lambda=4\) with interpolation, while ELFIC (Low Rate) covers only 0.15--0.56\,bpp. Overall, our method offers more configurations and continuous rate adaptation across five widths in a single model.

\noindent\textbf{Complexity and parameter efficiency.}
Our models achieve 25/37.5/50/75/100\% complexity with 6.34/12.68/15.84\,M parameters (E2E/Hyper/CKBD). ELFIC attains 7.66/35.82/74.06/100\% but costs 66.99\,M: it slims \(h_s\) and \(g_s\) yet zero-pads unused \(h_s\) features to keep full width for the channel-wise autoregressive context model \cite{minnen2020channel}, replaces GDN with NFFT, and widens \(y\) to 320 channels. CBANet-HP (6.39\,M, 25/50/100\%) is comparable to our E2E-192 (6.34\,M) but cannot interpolate; under the hyperprior backbone, our Hyper-192 achieves better RD than CBANet-HP with a moderate parameter increase. EVC (17.47\,M, 25/50/75/100\%) uses residual blocks, depth-wise separable convolutions, and mask decay that prunes inter-operator channels below the predefined widths. Overall, our method offers finer complexity control and better parameter efficiency.

\begin{table}[t!]
\centering
\caption{Summary of traditional codecs on Kodak. BPP and PSNR ranges are
computed over all evaluated quality points. Enc./Dec. denote the
average en-/de-coding time.}
\label{tab:traditional_summary}
\renewcommand{\arraystretch}{1.1}
\setlength{\tabcolsep}{3pt}
\begin{tabular}{lcccc}
\hline \hline
Codec & BPP & PSNR (dB) & Enc. (ms) & Dec. (ms) \\
\hline
JPEG       & 0.17--2.12 & 21.48--37.33 & 2.22 & 2.28 \\
JPEG2000   & 0.12--2.00 & 26.30--38.69 & 230.19 & 237.77 \\
BPG        & 0.10--2.12 & 27.14--42.12 & 470.27 & 168.11 \\
VTM-intra  & 0.11--2.11 & 28.43--41.17 & 140985.54 & 269.41 \\
\hline \hline
\end{tabular}
\end{table}

\noindent\textbf{Latency.}
Our CKBD-192 is faster than ELFIC across all widths (27.83--70.66\,ms encoding, 31.76--86.41\,ms decoding), and E2E-192/Hyper-192 are faster still. ELFIC slims \(g_s/h_s\), but its encoder zero-pads unused \(h_s\) channels before the channel-wise context model and keeps \(y\) full-width (320 channels). Since entropy coding dominates latency, its encoding is nearly constant (165.89\,ms) and decoding remains high (174.93--202.31\,ms), largely independent of complexity. CBANet-HP has a constant encoder (35.87\,ms) and extends the decoder by selecting among parallel decoder branches with weighted-sum outputs, so its decoding scales with complexity (54.03--58.32\,ms). EVC is fastest because it uses a custom multi-threaded C++ rANS entropy coder instead of CompressAI, minimizing Python/C++ overhead; its encoding grows only from 41.95 to 55.08\,ms as \(N_{\mathrm{enc}}\) increases from 1 to 4, while decoding stays at 34.12\,ms. With a CompressAI entropy coder, EVC would be close to our entropy coding latency (40.10/54.37\,ms in Table~\ref{tab:vbr_latency}), as it also uses a Gaussian conditional model. Overall, our VBR models offer a better trade-off among RD performance, parameter efficiency, and latency flexibility, while supporting continuous rate adaptation across multiple widths with a single model. The MS-SSIM results show the same trend as PSNR.

\subsection{Comparison with Traditional Codecs}
\label{subsec:traditional_codecs}

We compare our method with four traditional image and intra-frame compression codecs: JPEG, JPEG2000, BPG, and VTM~19.2 intra. All codecs are evaluated on the Kodak dataset, and the quality parameters are selected to cover a wide bitrate range, as detailed in the supplementary material.

Table~\ref{tab:traditional_summary} summarizes the BPP and PSNR ranges as well as the average encoding/decoding latency of each traditional codec. The corresponding rate--distortion curves are shown in Fig.~\ref{fig:psnr_compare}. Among the traditional codecs, VTM achieves the best rate--distortion performance, but at a high computational cost. As reported in Table~\ref{tab:traditional_summary}, its average encoding latency is substantially larger than that of JPEG, JPEG2000, and BPG, e.g., about 141~s per image for VTM. Our method achieves clear BD-rate savings over JPEG and JPEG2000, and remains competitive with BPG, while additionally offering flexible complexity and continuous rate adaptation. Traditional codecs run purely on the CPU, whereas LIC methods run transform networks on the GPU and entropy coding on the CPU.